\documentclass[final,a4paper,10pt]{article} 
\usepackage{amsmath}
\usepackage{amsthm,amssymb,amsfonts,amsxtra}
\usepackage{stmaryrd}
\usepackage[mathcal,mathscr]{euscript}
\usepackage[english]{babel}
\usepackage{times}
\usepackage{a4wide}
\usepackage[all]{xy}
\usepackage{fancyvrb}
\usepackage{afterpage}

\makeatletter
\newif\if@restonecol
\makeatother

\usepackage[boxed,titlenotnumbered,figure,commentsnumbered]{algorithm2e}

\usepackage{hyperref}

\theoremstyle{plain}
\newtheorem{theorem}{Theorem}[section]
\newtheorem{lemma}[theorem]{Lemma}
\newtheorem{corollary}[theorem]{Corollary}

\theoremstyle{definition}

\newtheorem{example}[theorem]{Example}

\theoremstyle{remark}

\numberwithin{equation}{section}

\theoremstyle{remark}

\newcommand{\comment} [1]{}

\def\ok#1{\mbox{\raisebox{0ex}[1ex][1ex]{$#1$}}}
\def \tuple#1{\langle #1 \rangle}

\newcommand{\mud}{{\mu^{\mathrm{d}}}}
\newcommand{\sra}{{\shortrightarrow}}

\newcommand{\CTL}{\ensuremath{\mathrm{CTL}}}
\newcommand{\LTL}{\ensuremath{\mathrm{LTL}}}
\newcommand{\CTLS}{\ensuremath{\mathrm{CTL\!}^*}}
\newcommand{\ACTLS}{\ensuremath{\mathrm{ACTL\!}^*}}
\newcommand{\ECTLS}{\ensuremath{\mathrm{ECTL\!}^*}}

\newcommand{\id}{\mathrm{id}}

\newcommand{\ud}{\mbox{\raisebox{0ex}[1ex][1ex]{$\:\stackrel{{\scriptscriptstyle
\mathrm{def}}}{=}\:$}}}

\newcommand{\ra}{\rightarrow}
\newcommand{\tle}{\trianglelefteq}
\newcommand{\vtl}{\vartriangleleft}
\newcommand{\Lra}{\Leftrightarrow}
\newcommand{\Ra}{\Rightarrow}

\newcommand{\cD}{{\mathcal{D}}}
\newcommand{\cK}{{\mathcal{K}}}

\newcommand{\cS}{{\mathcal{S}}}

\newcommand{\cB}{{\mathcal{B}}}

\newcommand{\cL}{{\mathcal{L}}}
\newcommand{\cX}{{\mathcal{X}}}

\newcommand{\cC}{{\mathcal{C}}}
\newcommand{\cP}{{\mathcal{P}}}

\newcommand{\Psim}{\ensuremath{P_{\mathrm{sim}}}}

\newcommand{\bZ}{\mathbb{Z}}

\newcommand{\It}{\mathit{It}}

\DeclareMathOperator{\Clv}{Cl_\cup}
\DeclareMathOperator{\Clc}{Cl_\cap}

\DeclareMathOperator{\ppRel}{\mathrm{pre}\mathit{PrevRel}}

\DeclareMathOperator{\AP}{{\!\mathit{AP}\!}}

\DeclareMathOperator{\pre}{pre}
\DeclareMathOperator{\pres}{pre_\shortrightarrow}
\DeclareMathOperator{\arr}{arr}
\DeclareMathOperator{\rem}{rem}
\DeclareMathOperator{\posts}{post_\shortrightarrow}

\DeclareMathOperator{\post}{post}
\DeclareMathOperator{\uco}{uco}

\DeclareMathOperator{\gfp}{gfp}

\DeclareMathOperator{\img}{img}

\DeclareMathOperator{\Part}{Part}

\DeclareMathOperator{\ucod}{uco^{\mathrm{d}}}

\DeclareMathOperator{\Rel}{\mathit{Rel}}

\DeclareMathOperator{\pr}{par}

\DeclareMathOperator{\Split}{\mathit{Split}}
\DeclareMathOperator{\splitt}{split}

\DeclareMathOperator{\SA}{SA}
\DeclareMathOperator{\BasicSA}{BasicSA}
\DeclareMathOperator{\RefinedSA}{RefinedSA}
\DeclareMathOperator{\HHK}{HHK}

\DeclareMathOperator{\parent}{parent}

\DeclareMathOperator{\Sim}{\mathit{Sim}}
\DeclareMathOperator{\prevSim}{\mathit{prevSim}}
\DeclareMathOperator{\Inv}{Inv}
\DeclareMathOperator{\Remove}{\mathit{Remove}}

\DefineVerbatimEnvironment%
{code}{Verbatim}{commandchars=\\\[\],
codes={\catcode`$=3\catcode`^=7\catcode`_=8},
frame=single}

\DefineVerbatimEnvironment%
{codenumber}{Verbatim}{commandchars=\\\[\],
codes={\catcode`$=3\catcode`^=7\catcode`_=8},
frame=lines, numbers=left}

\begin{document}

\title{\Large \bf An Efficient Simulation Algorithm
based on Abstract Interpretation}

\author{\normalsize {\sc Francesco Ranzato} ~~~~~ {\sc Francesco Tapparo}\\
\normalsize Dipartimento di Matematica Pura ed Applicata, University
of Padova, Italy\\
\normalsize \texttt{\{ranzato,tapparo\}$@$math.unipd.it}
}

\date{}
\maketitle
\pagestyle{plain}

\begin{abstract}
  A number of algorithms for computing the simulation
  preorder are available. Let $\Sigma$ denote the state space, $\sra$
  the transition relation and $P_{\mathrm{sim}}$ the partition of
  $\Sigma$ induced by simulation equivalence.  The algorithms by
  Henzinger, Henzinger, Kopke and by Bloom and Paige run in
  $O(|\Sigma||\sra|)$-time and, as far as time-complexity is
  concerned, they are the best available algorithms. However, these
  algorithms have the drawback of a space complexity that is more than
  quadratic in the size of the state space.  The algorithm by
  Gentilini, Piazza, Policriti~---~subsequently corrected by van
  Glabbeek and Ploeger~---~appears to provide the best compromise
  between time and space complexity. Gentilini et al.'s algorithm runs
  in $O(|P_{\mathrm{sim}}|^2 |\sra|)$-time while the space complexity
  is in $O(|P_{\mathrm{sim}}|^2 + |\Sigma| \log |P_{\mathrm{sim}}|)$.
  We present here a new efficient simulation  algorithm
  that is obtained as a modification of Henzinger et al.'s algorithm
  and whose correctness is based on some techniques used in
  applications of abstract interpretation to model checking. Our
  algorithm runs in $O(|P_{\mathrm{sim}}||\sra|)$-time and
  $O(|P_{\mathrm{sim}}| |\Sigma|\log|\Sigma|)$-space. Thus, this
  algorithm improves the best known time bound while retaining an
  acceptable space complexity that is in general less than quadratic
  in the size of the state space.  An experimental evaluation showed
  good comparative results with respect to Henzinger, Henzinger and
  Kopke's algorithm.
\end{abstract}

\section{Introduction}

Abstraction techniques are widely used in model checking to hide some
properties of the concrete model in order to define a reduced abstract
model where to run the verification algorithm~\cite{bk08,cgp99}. 
Abstraction provides an effective solution to deal with
the state-explosion problem that arises in model checking systems with
parallel components~\cite{cgjlv01}.  The reduced abstract structure is
required at least to weakly preserve a specification language $\cL$ of
interest: if a formula $\varphi\in\cL$ is satisfied by the reduced
abstract model then $\varphi$ must hold on the original
unabstracted model as well.  Ideally, the reduced model should be
strongly preserving w.r.t.\ $\cL$: $\varphi\in\cL$ holds on the
concrete model if and only if $\varphi$ is satisfied by the reduced
abstract model. One common approach for abstracting a model consists
in defining a logical equivalence or preorder on system
states that weakly/strongly preserves a given temporal language. 
Moreover, this equivalence or preorder often arises as 
a behavioural relation in the context of process calculi~\cite{cs01}.
Two
well-known examples are bisimulation equivalence that strongly
preserves expressive logics such as $\CTLS$ and the full
$\mu$-calculus~\cite{bcg88} and the simulation preorder that ensures
weak preservation of universal and existential fragments of the
$\mu$-calculus like $\ACTLS$ and $\ECTLS$ as well as of linear-time
languages like $\LTL$~\cite{gl94,loi95}. Simulation equivalence,
namely the equivalence relation obtained as symmetric reduction of the
simulation preorder, is particularly interesting because
it can provide a  significantly better state space reduction
than bisimulation equivalence while retaining the ability of
strongly preserving expressive temporal languages like $\ACTLS$.

\paragraph*{State of the Art.} 
It is known that computing simulation is harder than
computing bisimulation~\cite{km02}.  Let
$\cK=\tuple{\Sigma,\sra,\ell}$ denote a Kripke structure on the state
space $\Sigma$, with transition relation $\sra$ and labeling function
$\ell\!:\!\Sigma\!\ra\! \wp(\AP)$, for a given set $\AP$ of atomic
propositions. Bisimulation equivalence can be computed by the
well-known Paige and Tarjan's~\cite{pt87} algorithm that runs in
$O(|\sra|\log|\Sigma|)$-time.  A number of algorithms for computing
simulation equivalence exist, the most well known are by Henzinger,
Henzinger and Kopke~\cite{hhk95}, Bloom and Paige~\cite{bp95}, Bustan
and Grumberg~\cite{bg03}, Tan and Cleaveland~\cite{TC01} and
Gentilini, Piazza and Policriti~\cite{gpp03}, this latter subsequently
corrected by van Glabbeek and Ploeger~\cite{GP08}.  
The algorithms by Henzinger, Henzinger, Kopke and by
Bloom and Paige run in $O(|\Sigma||\sra|)$-time and, as far as
time-complexity is concerned, they are the best available
algorithms. However,
both these algorithms have the drawback of a space complexity
that is bounded from below by $\Omega(|\Sigma|^2)$. This is due to the fact
that the simulation preorder is computed in an explicit way, i.e., 
for any state $s\in \Sigma$, the set of states
that simulate $s$ is explicitly given as output. 
 This quadratic
lower bound in the size of the state space is clearly a critical
issue in model checking. There is therefore a strong
motivation for designing simulation algorithms that are
less demanding on space requirements.  Bustan and
Grumberg~\cite{bg03} provide a first solution in this direction.  Let
$P_{\mathrm{sim}}$ denote the partition corresponding to simulation
equivalence on $\cK$ so that $|P_{\mathrm{sim}}|$ is the number of
simulation equivalence classes.  Then, Bustan and Grumberg's algorithm
has a space complexity in $O(|P_{\mathrm{sim}}|^2 + |\Sigma| \log
|P_{\mathrm{sim}}|)$, although the time complexity in
$O(|P_{\mathrm{sim}}|^4(|\sra|+|P_{\mathrm{sim}}|^2) +
|P_{\mathrm{sim}}|^2|\Sigma|(|\Sigma|+|P_{\mathrm{sim}}|^2|))$ remains
a serious drawback.  The simulation algorithm by Tan and
Cleaveland~\cite{TC01} simultaneously computes also the 
state partition $P_{\mathrm{bis}}$ corresponding to bisimulation
equivalence.   Under 
the simplifying assumption of dealing with a total transition relation, this
procedure has a 
time complexity  in $O(|\sra|(|P_{\mathrm{bis}}| + \log |\Sigma|))$ 
and a space complexity in $O(|\sra| + |P_{\mathrm{bis}}|^2 +
|\Sigma|\log|P_{\mathrm{bis}}|)$ (the latter factor
$|\Sigma|\log|P_{\mathrm{bis}}|$ does not appear in \cite{TC01} and
takes into account the relation that maps each state into its
bisimulation equivalence class). 
The algorithm by Gentilini, Piazza and
Policriti~\cite{gpp03} appears to provide the best compromise between time
and space complexity.  Gentilini et al.'s
algorithm runs in $O(|P_{\mathrm{sim}}|^2 |\sra|)$-time, namely it remarkably
improves on Bustan and Grumberg's algorithm and is not directly
comparable
with Tan and Cleaveland's algorithm, while the space
complexity $O(|P_{\mathrm{sim}}|^2 + |\Sigma| \log
|P_{\mathrm{sim}}|)$ is the same of Bustan and Grumberg's
algorithm and improves on Tan and
Cleaveland's algorithm.  Moreover, Gentilini et al.\
show experimentally that in most cases their procedure improves
on Tan and Cleaveland's algorithm both in time and space.

\paragraph*{Main Contributions.} 
This work presents a new efficient simulation  algorithm, called 
$\SA$,
that runs in $O(|P_{\mathrm{sim}}||\sra|)$-time and
$O(|P_{\mathrm{sim}}| |\Sigma| \log|\Sigma|)$-space. Thus, while
retaining an acceptable space
complexity  that is in general less than quadratic
in the size of the state space, our algorithm improves the best known
time
bound. 
\\
\indent 
Let us recall that a relation $R$ between states is a
simulation if for any $s,s'\in \Sigma$ such that $(s,s')\in R$,
$\ell(s) = \ell (s')$ and for any $t\in \Sigma $ such that $\ok{s \sra
  t}$, there exists $t'\in \Sigma $ such that $\ok{s'\sra t'}$ and
$(t,t')\in R$. Then, $s'$ simulates $s$, namely the pair $(s,s')$ belongs 
to the simulation preorder $R_{\mathrm{sim}}$, if there exists a
simulation relation $R$ such $(s,s')\in R$. Also, $s$ and $s'$ are
simulation equivalent, namely they belong to the same
block of the simulation partition $P_{\mathrm{sim}}$, 
if $s'$ simulates $s$ and vice versa. 
\\
\indent
Our simulation  algorithm $\SA$ is designed as a
modification of Henzinger, Henzinger and Kopke's~\cite{hhk95}
algorithm, here denoted by $\HHK$.  The space complexity of 
$\HHK$ is in $O(|\Sigma|^2\log |\Sigma|)$. This is a consequence of the fact
that  $\HHK$ computes explicitly the simulation preorder, namely it 
maintains for any state $s\in \Sigma$ a set of states
$\Sim(s)\subseteq \Sigma$, called the simulator set of $s$, which
stores states that are currently candidates for simulating $s$.  
Our algorithm $\SA$ computes 
instead a symbolic representation of the simulation preorder, namely
it maintains:
(i) a partition $P$ of the state space $\Sigma$ that is always
coarser than the final simulation partition $P_{\mathrm{sim}}$ and (ii) a relation $\Rel
\subseteq P\times P$ on the current partition $P$ that encodes the
simulation relation between blocks of simulation equivalent
states. This symbolic representation is the key  both for obtaining the 
$O(|\Psim||\sra|)$ time bound and for limiting the space
complexity of $\SA$ in
$O(|P_{\mathrm{sim}}| |\Sigma|\log |\Sigma|)$, so that memory requirements may
be lower than quadratic in the size of the state space.
\\
\indent
The basic idea of our approach is to investigate whether the
logical structure of the $\HHK$ algorithm may be preserved
by replacing the family of sets of states $\cS=\{\Sim(s)\}_{s\in
  \Sigma}$ with the following state
partition $P$ induced by $\cS$: two states 
$s_1$ and $s_2$ are equivalent in $P$ iff for all
$s\in \Sigma$, $s_1 \in \Sim(s)\Lra s_2 \in \Sim(s)$.  
Additionally, we store and maintain a preorder relation
$\Rel\subseteq P\times P$ on the partition $P$ that gives rise to a
so-called partition-relation pair $\tuple{P,\Rel}$.  The logical
meaning of this data structure is that 
if $B,C\in P$ and $(B,C)\in \Rel$ then any state in $C$ is currently
candidate to simulate each state in $B$, while two states $s_1$ and $s_2$ 
in the same block $B$ are currently candidates to be
simulation equivalent. 
Hence, a partition-relation pair
$\tuple{P,\Rel}$ represents the current approximation of the
simulation preorder and in particular $P$ represents the current
approximation of simulation equivalence. 
It turns out that the information encoded by a
partition-relation pair is enough for preserving the logical structure
of $\HHK$. In fact, analogously to the stepwise design of
the $\HHK$ procedure, this approach leads us to design a basic
procedure $\BasicSA$ based on partition-relation pairs which is then
refined twice in order to obtain  
the final simulation algorithm $\SA$. The correctness of $\SA$
is proved w.r.t.\ the basic algorithm $\BasicSA$ and relies on
abstract interpretation techniques~\cite{CC77,CC79}. 
More specifically, we exploit some previous
results~\cite{rt07} 
that show how standard strong preservation of temporal
languages in abstract Kripke structures can be generalized by abstract
interpretation and cast as a so-called completeness property of
abstract domains. On the other hand,
the simulation algorithm $\SA$ is designed  as 
an efficient implementation of the basic procedure $\BasicSA$ where 
the symbolic representation based on partition-relation pairs allows
us to replace the size $|\Sigma|$ of the state space 
in the time and space bounds of $\HHK$ with the size
$|P_{\mathrm{sim}}|$ of the
simulation partition  in the corresponding bounds
for $\SA$. 
\\
\indent
Both $\HHK$ and $\SA$ have been implemented in C++. This practical
evaluation considered benchmarks from 
the VLTS (Very Large Transition Systems) 
suite~\cite{vasy}  and some publicly available 
Esterel programs. The experimental results  
showed that $\SA$  outperforms $\HHK$.

\section{Background}\label{background}

\subsection{Preliminaries}

\paragraph{Notations.} 
Let $X$ and $Y$ be sets. 
If $S\subseteq X$ and $X$ is understood as a universe set then 
$\neg S = X\smallsetminus S$. 
If $f:X\ra Y$ then the
image of $f$ is denoted by $\img(f)=\{f(x)\in Y~|~ x\in X\}$. 
When writing a set $S$ of
subsets of a given set of integers, e.g.\ a partition, $S$ is often written in a
compact form like $\{1, 12, 13\}$ or $\{[1], [12], [13]\}$ that stands
for $\{\{1\}, \{1,2\}, \{1,3\}\}$. 
If $R\subseteq X\times X$ is any
relation then $R^*\subseteq X\times X$ denotes the reflexive
and transitive closure of $R$. Also, if $x\in X$ then $R(x)\ud \{x'\in
X~|~ (x,x')\in R\}$.

\paragraph*{Orders.} 
Let $\tuple{Q,\leq}$ be a poset, that may also be denoted by $Q_\leq$.
We use the symbol $\sqsubseteq$ to denote
pointwise ordering between functions: If $X$ is any set and $f,g:X \ra
Q$ then $f\sqsubseteq g$ if for all $x\in X$, $f(x)\leq g(x)$. 
If $S\subseteq Q$ then $\max (S) \ud \{ x\in S~|~ \forall y\in S.\; x\leq y \Ra x
= y\}$ denotes the set of maximal elements of $S$ in $Q$. 
A complete lattice $C_\leq$ is also denoted
by $\tuple{C,\leq,\vee,\wedge,\top,\bot}$ where $\vee$, $\wedge$,
$\top$ and $\bot$ denote, respectively, lub, glb, greatest element and
least element in $C$.  A function $f:C\ra D$ between complete lattices
is additive when $f$ preserves least upper bounds.  
Let us recall that a reflexive and transitive 
relation $R\subseteq X\times X$ on a set $X$ is called a
preorder on $X$.

\paragraph*{Partitions.}
A partition $P$ of a set $\Sigma$ is a set of nonempty subsets of
$\Sigma$, called blocks, that are pairwise disjoint and whose union
gives $\Sigma$.  $\Part(\Sigma)$ denotes the set of partitions of
$\Sigma$.  If $P\in \Part(\Sigma)$ and $s\in \Sigma$ then $P(s)$
denotes the block of $P$ that contains $s$. $\Part(\Sigma)$ is endowed
with the following standard partial order $\preceq$: $P_1 \preceq
P_2$, i.e.\ $P_2$ is coarser than $P_1$ (or $P_1$ refines $P_2$) iff
$\forall B\in P_1. \exists B' \in P_2 .\: B \subseteq B'$.  If
$P_1,P_2\in \Part(\Sigma)$, $P_1\preceq P_2$ and $B\in P_1$ then
$\parent_{P_2}(B)$ (when clear from the context the subscript $P_2$
may be omitted) denotes the unique block in $P_2$ that contains $B$.
For a given nonempty subset $S\subseteq \Sigma$ called splitter, we
denote by $\Split(P,S)$ the partition obtained from $P$ by replacing
each block $B\in P$ with the nonempty sets $B\cap S$ and
$B\smallsetminus S$, where we also allow no splitting, namely
$\Split(P,S)=P$ (this happens exactly when $S$ is a union of some
blocks of $P$).

\paragraph*{Kripke Structures.} 
A transition system $(\Sigma ,\sra)$ consists of a 
set $\Sigma$ of states and a transition relation $\sra 
\subseteq \Sigma \times
\Sigma$. 
The relation $\sra$ is total when
for any $s\in \Sigma $ there exists some $t\in \Sigma $ such
that $s\sra t$. 
The predecessor/successor 
transformers $\pres,\post_{\sra}:\wp(\Sigma)\ra \wp(\Sigma)$ (when
clear from the context the subscript $\sra$ may be omitted) are
defined as usual:
\begin{itemize}
\item[--] $\pre_\sra (Y) \ud \{ a\in \Sigma ~|~\exists b\in Y.\; a \sra
b\}$;
\item[--] $\post_{\sra}(Y) \ud \{ b \in \Sigma~|~ \exists a\in Y.\;
a\sra b\}$.
\end{itemize} 
Let us remark that $\pre_\sra$ and $\post_\sra$ are additive operators on
the complete lattice $\wp(\Sigma)_\subseteq$.
If $S_1,S_2\subseteq \Sigma$ then $S_1
\sra^{\exists\exists} S_2$ iff there exist $s_1\in S_1$ and $s_2\in
S_2$ such that $s_1 \sra s_2$.  

Given a set $\mathit{AP}$ of atomic propositions
(of some specification language), a Kripke structure $\cK= (\Sigma ,\sra,\ell)$ over
$\mathit{AP}$ consists of a transition system $(\Sigma ,\sra)$
together with a state labeling function $\ell:\Sigma \ra
\wp(\mathit{AP})$. A Kripke structure is called total when its
transition relation is total. We use the following notation: 
for any $s\in \Sigma$, $[s]_\ell \ud \{s'\in
\Sigma~|~ \ell(s)=\ell(s')\}$ denotes the equivalence class of a state
$s$ w.r.t.\ the labeling $\ell$, while $P_\ell \ud
\{[s]_\ell~|~ s\in \Sigma\}\in\Part(\Sigma)$ is the partition induced by $\ell$.

\subsection{Simulation Preorder and Equivalence}\label{spe}

Recall that a relation 
$R\subseteq \Sigma \times \Sigma $ is a
simulation on a Kripke structure $\cK=(\Sigma,\sra, \ell)$ over a set
$\AP$ of atomic propositions 
if for any $s,s'\in \Sigma$ such that $(s,s')\in R$:
\begin{itemize} 
\item[{\rm (a)}] $\ell(s) = \ell (s')$; 
\item[{\rm (b)}] For
any $t\in \Sigma $ such that $\ok{s \sra t}$, there exists $t'\in \Sigma $
such that $\ok{s'\sra t'}$
and $(t,t')\in R$.
\end{itemize}

\noindent
If $(s,s')\in R$ then we say that $s'$ simulates $s$. 
The empty
relation is a simulation and simulation relations are closed under
union, so that the 
largest simulation relation exists. It turns out that the largest
simulation is a preorder relation called simulation preorder (on
$\cK$) and
denoted by $R_{\mathrm{sim}}$. 
Simulation equivalence 
$\sim_{\mathrm{sim}}\,\subseteq
\Sigma \times \Sigma $ is the symmetric reduction of 
$R_{\mathrm{sim}}$, namely $\sim_{\mathrm{sim}} = R_{\mathrm{sim}}
\cap R_{\mathrm{sim}}^{-1}$. 
$P_{\mathrm{sim}}\in \Part(\Sigma)$ denotes
the partition corresponding to $\sim_{\mathrm{sim}}$ and is called
simulation partition.

It is a well known result in model checking \cite{dgg93,gl94,loi95} that
the reduction  of $\cK$ w.r.t.\ simulation equivalence
$\sim_{\mathrm{sim}}$ allows us to define an
abstract Kripke structure
$\mathcal{A}_{\mathrm{sim}}=\tuple{P_{\mathrm{sim}},\sra^{\exists\exists}, \ell^\exists}$ 
that strongly preserves the temporal
language $\ACTLS$, where: $P_{\mathrm{sim}}$  is the abstract
state space, $\sra^{\exists\exists}$ is the 
abstract transition relation between simulation equivalence classes,
while for any block $B\in P_{\mathrm{sim}}$, $\ell^\exists(B)\ud \ell(s)$ for
any representative $s\in B$. It turns out that
$\mathcal{A}_{\mathrm{sim}}$ strongly preserves $\ACTLS$, i.e., for any
$\varphi\in \ACTLS$, $B\in P_{\mathrm{sim}}$ and $s\in B$, we have
that $s\models^\cK \varphi$ if and only if
$B\models^{\mathcal{A}_{\mathrm{sim}}} \varphi$.

\subsection{Abstract Interpretation}

\paragraph*{Abstract Domains as Closures.} 
In standard abstract interpretation, abstract domains can be
equivalently specified either by Galois connections/insertions or by
(upper) closure operators (uco's)~\cite{CC79}.  These two approaches
are equivalent, modulo isomorphic representations of domain's objects.
We follow here the closure operator approach: this has the advantage
of being independent from the representation of domain's objects and
is therefore appropriate for reasoning on abstract domains
independently from their representation.  

Given a state space
$\Sigma$, the complete lattice $\wp(\Sigma)_\subseteq$ plays
the role of concrete domain.  Let us recall that an operator
$\mu:\wp(\Sigma)\ra \wp(\Sigma)$ is a uco on $\wp(\Sigma)$, that is an
abstract domain of $\wp(\Sigma)$, when $\mu$ is monotone, idempotent
and extensive (viz., $X \subseteq \mu (X)$).  It is well known that
the set $\uco(\wp(\Sigma))$ of all uco's on $\wp(\Sigma)$, endowed
with the pointwise ordering $\sqsubseteq$, gives rise to the complete
lattice $\tuple{\uco(\wp(\Sigma)), \sqsubseteq,\sqcup,\sqcap,\lambda
  X.\Sigma, \id}$ of all the abstract domains
of $\wp(\Sigma)$.  The pointwise ordering $\sqsubseteq$ on $\uco
(\wp(\Sigma))$ is the standard order for comparing abstract domains
with regard to their precision: $\mu_1 \sqsubseteq \mu_2$ means that
the domain $\mu_1$ is a more precise abstraction of $\wp(\Sigma)$ than
$\mu_2$, or, equivalently, that the abstract domain $\mu_1$ is a
refinement of $\mu_2$.

A closure $\mu \in\uco(\wp(\Sigma))$ is uniquely determined by its
image $\img(\mu)$, which coincides with its set of fixpoints, as
follows: $\mu= \lambda Y.\cap \{ X\in \img(\mu)~|~ Y\subseteq
X\}$. Also, a set of subsets $\cX\subseteq \wp(\Sigma)$ is the image
of some closure operator $\mu_\cX\in\uco(\wp(\Sigma))$ iff $\cX$ is a Moore-family of $\wp(\Sigma)$, i.e.,
$\cX=\Clc (\cX)\ud \{\cap S~|~ S\subseteq \cX\}$ (where $\cap
\varnothing=\Sigma \in \Clc (\cX)$). In other terms, $\cX$ is a
Moore-family (or Moore-closed) when $\cX$ is closed
under arbitrary intersections.  In this case, $\mu_\cX=\lambda Y.\cap
\{ X\in \cX~|~ Y\subseteq X\}$ is the corresponding closure operator.
For any $\cX\subseteq \wp(\Sigma)$, $\Clc (\cX)$ is called the
Moore-closure of $\cX$, i.e., $\Clc (\cX)$ is the least set of subsets
of $\Sigma$ which contains all the subsets in $\cX$ and is
Moore-closed. Moreover, it turns out that for any $\mu \in
\uco(\wp(\Sigma))$ and any Moore-family $\cX\subseteq \wp(\Sigma)$,
$\mu_{\img(\mu)} = \mu$ and $\img(\mu_\cX)=\cX$. Thus, closure
operators on $\wp(\Sigma)$ are in bijection with Moore-families of
$\wp(\Sigma)$. This allows us to consider a closure operator $\mu\in
\uco(\wp(\Sigma))$ both as a function $\mu:\wp(\Sigma)\ra \wp(\Sigma)$
and as a Moore-family $\img(\mu)\subseteq \wp(\Sigma)$.  This is
particularly useful and does not give rise to ambiguity since one can
distinguish the use of a closure $\mu$ as function or set according to
the context.

\paragraph*{Abstract Domains and Partitions.}  As shown in \cite{rt07},
it turns out that partitions can be viewed as particular abstract
domains.  Let us recall here that any abstract domain $\mu\in
\uco(\wp(\Sigma))$ induces a partition $\pr(\mu)\in
\Part(\Sigma)$ that corresponds to the following equivalence relation 
$\equiv_\mu$ on $\Sigma$:
$$x\equiv_\mu y
\;\text{~iff~}\; \mu(\{x\}) = \mu(\{y\}).$$

\begin{example}\label{simple}
Let $\Sigma =\{1,2,3,4\}$ and consider the following 
abstract domains in $\uco(\wp(\Sigma))$ 
that are given as intersection-closed 
subsets of $\wp(\Sigma)$: 
$\mu = \{\varnothing,3,4,12,34,1234\}$, 
$\mu' =\{\varnothing,3,4,12,1234\}$, 
$\mu'' = \{ 12, 123, 124, 1234\}$. These abstract domains 
all  induce the same
partition $P= \{[12],[3],[4]\}\in \Part(\Sigma)$. For example, 
$\mu''(\{1\}) =\mu''(\{2\})= \{1,2\},~\mu'' (\{3\})= \{1,2,3\}$,
$\mu''(\{4\}) = \{1,2,4\}$ so that $\pr(\mu'')=P$. 
 \qed
\end{example}

\paragraph*{Forward Completeness.}
Let us consider an abstract domain
$\mu\in \uco(\wp(\Sigma)_\subseteq)$, 
a concrete semantic function $f:\wp(\Sigma)\ra \wp(\Sigma)$
and a corresponding abstract
semantic function $f^\sharp:\mu \ra \mu$ (for simplicity of notation, 
we consider 1-ary functions). 
It is well known that the abstract interpretation 
$\ok{\tuple{\mu,f^\sharp}}$ is sound
when $\ok{f \circ \mu  \sqsubseteq 
  f^\sharp \circ \mu}$ holds: this means that a concrete computation
$f(\mu(X))$ on an abstract object $\mu(X)$ 
is correctly approximated in $\mu$ by $\ok{f^\sharp(\mu(X))}$,
that is, $f(\mu(X)) \subseteq  \ok{f^\sharp(\mu(X))}$. 
Forward completeness corresponds to require
the following strengthening of soundness: $\ok{\tuple{\mu,f^\sharp}}$ is
forward complete when $\ok{f\circ \mu = 
f^\sharp \circ \mu}$: The intuition here is that the abstract function
$\ok{f^\sharp}$ is able to mimic $f$
on the abstract domain $\mu$ with no loss of precision.  This is called
forward completeness because a dual and more standard notion of backward completeness 
may also be considered (see e.g.\
\cite{gq01}). 

\begin{example}
As a toy
example, let us consider the following abstract domain $\mathit{Sign}$
for representing the sign of an integer variable: $\mathit{Sign} =
\{\varnothing, \ok{\bZ_{\scriptscriptstyle{\leq 0}}}, 0,
\ok{\bZ_{\scriptscriptstyle{\geq 0}}}, \bZ\}\in
\uco(\wp(\bZ)_\subseteq )$. The concrete pointwise
addition $+:\wp(\bZ)\times \wp(\bZ) \ra \wp(\bZ)$ on sets
of integers, that is 
$X+Y \ok{\ud} \{x+y~|~ x\in X,\, y\in Y\}$, is approximated in $\mathit{Sign}$ by the
abstract addition $\ok{+^\mathit{Sign}}:\mathit{Sign}\times
\mathit{Sign}\ra \mathit{Sign}$ that is
defined as expected by the following table:
\[
\begin{array}{|l||l|l|l|l|l|}
\hline 
\!+^\mathit{Sign}\!\! & \varnothing & \ok{\bZ_{\scriptscriptstyle{\leq
      0}}} & 0 & \ok{\bZ_{\scriptscriptstyle{\geq 0}}} & \bZ \\
\hline \hline
\varnothing & \varnothing &  \varnothing &  \varnothing &  \varnothing &  \varnothing 
\\ \hline
\ok{\bZ_{\scriptscriptstyle{\leq 0}}} & \varnothing &
\ok{\bZ_{\scriptscriptstyle{\leq 0}}} &
\ok{\bZ_{\scriptscriptstyle{\leq 0}}} & \bZ & \bZ  
\\ \hline
0 & \varnothing & \ok{\bZ_{\scriptscriptstyle{\leq
      0}}} & 0 & \ok{\bZ_{\scriptscriptstyle{\geq 0}}} & \bZ
\\ \hline
\ok{\bZ_{\scriptscriptstyle{\geq 0}}} & \varnothing & \bZ &
\ok{\bZ_{\scriptscriptstyle{\geq 0}}} &
\ok{\bZ_{\scriptscriptstyle{\geq 0}}} & \bZ  
\\ \hline
\bZ & \varnothing &\bZ &\bZ &\bZ &\bZ 
\\ \hline
\end{array}
\]

\noindent
It turns out that $\tuple{\mathit{Sign},\ok{+^\mathit{Sign}}}$ 
is forward complete, i.e., for any $a_1,a_2 \in
\mathit{Sign}$, $a_1 + a_2 = 
a_1 \:\ok{+^\mathit{Sign}}\: a_2$. \qed 
\end{example}

It turns out that the possibility of defining a forward complete
abstract interpretation on a given abstract domain $\mu$ does not depend
on the choice of the abstract function $f^\sharp$  but depends only on
the abstract domain $\mu$. 
This
means that if $\ok{\tuple{\mu,f^\sharp}}$ is forward complete  then 
the abstract function $\ok{f^\sharp}$ indeed coincides
with the best correct approximation $\mu \circ f$ of the concrete function $f$ on the
abstract domain $\mu$. 
Hence, for any abstract domain $\mu$ and abstract function $\ok{f^\sharp}$, 
it turns out that $\ok{\tuple{\mu,f^\sharp}}$ is forward complete if and
only if $\tuple{\mu,\mu\circ f}$ is forward complete. 
This allows us to define the notion of forward completeness
independently of abstract functions as follows: an
abstract domain $\mu\in\uco(\wp(\Sigma))$ is forward complete for $f$
(or forward $f$-complete)
iff $f \circ \mu = \mu \circ f \circ \mu$. 
Let us remark that
$\mu$ is forward $f$-complete iff the image $\img(\mu)$ is closed
under applications of the concrete function $f$. If $F$ is a set of concrete functions then
$\mu$ is forward complete for $F$ when $\mu$ is forward complete for
all $f\in F$.

\paragraph*{Forward Complete Shells.}
It turns out \cite{gq01,rt07} that any abstract domain $\mu \in
\uco(\wp(\Sigma))$ can be refined to its forward $F$-complete shell,
namely to the most abstract domain that is forward complete for $F$
and refines $\mu$. This forward $F$-complete shell of $\mu$ 
is thus defined as 
$$\cS_F(\mu) \ud \sqcup \{ \rho \in \uco(\wp(\Sigma))~|~ \rho
\sqsubseteq \mu,\, \rho \text{~is forward $F$-complete}\}.$$

Forward complete shells admit a constructive
fixpoint characterization. 
Given $\mu \in \uco (\wp(\Sigma))$,  consider the operator
$F_\mu: \uco(\wp(\Sigma))\ra \uco(\wp(\Sigma))$ defined by 
$$F_\mu (\rho) \ud  \Clc (\mu \cup \{f(X)~|~f\in F,\: X\in \rho\}).$$ 
Thus, $F_\mu(\rho)$ refines the abstract domain $\mu$ by adding the
images of $\rho$ for all the functions in $F$. 
It turns out that $F_\mu$ is
monotone and therefore 
admits the greatest
fixpoint, denoted by $\gfp(F_\mu)$, 
which provides the forward $F$-complete shell of $\mu$: $\cS_F (\mu) =
\gfp (F_\mu )$.

\paragraph*{Disjunctive Abstract Domains.} 
An abstract domain $\mu \in \uco(\wp(\Sigma))$ is disjunctive (or
additive) when $\mu$ is additive
and this happens exactly when the image $\img(\mu)$ is closed
under arbitrary unions. Hence, a disjunctive abstract domain is
completely determined by the image of $\mu$ on singletons because for
any $X\subseteq \Sigma$, $\mu(X)=\cup_{x\in X} \mu(\{x\})$.  The
intuition is that a disjunctive abstract domain does not lose
precision in approximating concrete set unions.  We denote by
$\ucod(\wp(\Sigma))\subseteq \uco(\wp(\Sigma))$ the set of disjunctive
abstract domains.

Given
any abstract domain $\mu \in \uco(\wp(\Sigma))$, it turns
out~\cite{CC79,gr98} that $\mu$ can be refined to its disjunctive
completion $\mu^{\mathrm{d}}$: this is  the most abstract disjunctive
domain $\mu^{\mathrm{d}} \in \ucod(\wp(\Sigma))$ that refines $\mu$.  
The disjunctive completion $\mu^{\mathrm{d}}$ can
be obtained by closing the image $\img(\mu)$ under arbitrary unions,
namely $\img(\mu^{\mathrm{d}}) = \Clv(\img(\mu)) \ud \{\cup
\mathcal{S}~|~ \mathcal{S}\subseteq \img(\mu)\}$, where $\cup
\varnothing = \varnothing \in \Clv(\img(\mu))$.

It turns out that an abstract domain $\mu$ is disjunctive iff
$\mu$ is forward complete for arbitrary concrete set unions, namely,
$\mu$ is disjunctive iff for any $\{X_i\}_{i\in I} \subseteq
\wp(\Sigma)$, $\cup_{i\in I} \mu(X_i) = \mu (\cup_{i\in I}
\mu(X_i))$. Thus, when $\Sigma$ is finite, the disjunctive completion
$\mu^{\mathrm{d}}$ of $\mu$ coincides with the forward $\cup$-complete
shell $\cS_\cup(\mu)$ of $\mu$.
Also, since the predecessor transformer $\pres$
preserves set unions, it turns out that the forward complete shell
$\cS_{\cup,\pres}(\mu)$ for $\{\cup,\pres\}$ 
can be obtained by iteratively closing  the image of
$\mu$
under $\pres$ and then by taking the
disjunctive completion, i.e., 
$\cS_{\cup,\pres}(\mu) = \cS_\cup (\cS_{\pres} (\mu))$.

\begin{example}\label{simple2}
Let us consider the abstract domain $\mu = \{\varnothing,
3, 4, 12, 34, 1234\}$ in
Example~\ref{simple}. We have that $\mu$ is not disjunctive because
$12,3\in \mu$ while $12\cup 3=123\not \in \mu$.  
The disjunctive
completion $\mu^{\mathrm{d}}$ is obtained by closing $\mu$ under
unions: $\mu^{\mathrm{d}} = \{\varnothing,3,4, 12, 34, 123, 124,
1234\}$.   \qed
\end{example}

\paragraph*{Some Properties of Abstract Domains.} 

Let us summarize some easy properties of abstract domains
that will be used in later proofs. 

\begin{lemma}\label{prop}
Let $\mu \in \uco(\wp(\Sigma))$, $\rho \in \ucod(\wp(\Sigma))$,
$P,Q\in \Part(\Sigma)$ such that  $P\preceq
  \pr(\mu)$ and $Q\preceq \pr(\rho)$.
\begin{itemize}
\item[{\rm (i)}] For any $B\in P$, 
$\mu(B) = \mu(\parent_{\pr(\mu)}(B))$.

\item[{\rm (ii)}] For any $X\in \wp(\Sigma)$,
   $\mu(X) = \cup \{ B\in P~|~B \subseteq \mu(X)\}$.
\item[{\rm (iii)}] For any $X\in \wp(\Sigma)$, $\rho(X)= \cup \{\rho(B)~|~
  B\in Q,\: B\cap X\neq \varnothing\}$.  
\item[{\rm (iv)}] $\pr(\mu)=\pr(\mud)$. 
\end{itemize} 
\end{lemma}
\begin{proof}
(i)~~In general, by definition of $\pr(\mu)$, 
for any $C\in \pr(\mu)$ and $S\subseteq C$, $\mu(S)=\mu(C)$. Hence,
since $B\subseteq \parent_{\pr(\mu)}(B)$ we have that 
$\mu(B) = \mu(\parent_{\pr(\mu)}(B))$.\\
(ii)~~Clearly, $\mu(X) \supseteq \cup \{ B\in P~|~B \subseteq
\mu(X)\}$. On the other hand, given $z\in \mu(X)$, let $B_z\in P$ be the
block in $P$ that contains $z$. Then, $B_z \subseteq \mu(B_z)=\mu(\{z\}) \subseteq
\mu(X)$, so that $z\in \cup \{ B\in P~|~B \subseteq
\mu(X)\}$. \\
(iii)
\begin{align*}
\rho(X) & = \text{~~~~~[as $\rho$ is additive]}\\
\cup \{\rho(\{x\})~|~x\in X\} & = \text{~~~~~[as $Q\preceq \pr(\rho)$]}\\
\cup \{\rho(B_x)~|~x\in X,\; B_x\in Q,\; x\in B_x\} & = \\
\cup \{\rho(B)~|~
  B\in Q,\: B\cap X\neq \varnothing\}&.
\end{align*}
(iv)~~Since $\mud\sqsubseteq \mu$, we have that $\pr(\mud)\preceq
\pr(\mu)$. On the other hand, if $B\in \pr(\mu)$ then for all $x\in
B$, $\mud(\{x\})=\mu(\{x\})=\mu(B)$, so that $B\in \pr(\mud)$. 
\end{proof}

\section{Simulation Preorder as a Forward Complete Shell}

Ranzato and Tapparo \cite{rt07} showed how strong
preservation of specification languages in  standard abstract
models like abstract Kripke structures
can be generalized by abstract
interpretation and cast as a forward completeness property of generic abstract
domains that play the role of abstract models. 
We rely here on this framework in order to 
show that the simulation preorder can be characterized as a 
forward complete shell for set union and the predecessor
transformer. 
Let $\cK=(\Sigma,\sra,\ell)$ be a Kripke structure. Recall that the labeling
function $\ell$ induces the state partition $P_\ell = \{[s]_\ell ~|~
s\in \Sigma\}$. This partition can be made an abstract domain 
$\mu_\ell\in
\uco(\wp(\Sigma))$ by considering the Moore-closure of $P_\ell$ that
simply adds to $P_\ell$ the empty set and the whole state space, namely   
$\mu_\ell \ud \Clc(\{[s]_\ell ~|~ s\in \Sigma\})$. 
\begin{theorem}\label{fcs}
Let $\mu_\cK = \cS_{\cup,\pre}(\mu_\ell)$ be the
forward $\{\cup,\pre\}$-complete shell of $\mu_\ell$. Then,
$R_{\mathrm{sim}} =\{(s,s')\in \Sigma\times \Sigma~|~  s' \in
\mu_\cK(\{s\})$ and 
$P_{\mathrm{sim}} = \pr(\mu_\cK)$. 
\end{theorem}
\begin{proof}
Given a disjunctive abstract domain $\mu\in \ucod(\wp(\Sigma))$, define
  $R_\mu
\ud \{(s,s')\in\Sigma\times \Sigma~|~s'\in \mu(\{s\})\}$. 
  We prove the following three preliminary facts:
  \begin{itemize}
    
    \item[{\rm (1)}] $\mu$ is forward complete for $\pre$ iff $R_\mu$ satisfies 
      the following property: for any $s,t,s'\in\Sigma$ such that $s\ra t$ and
      $(s,s')\in R_\mu$ there exists $t'\in \Sigma$ such that $s'\ra
      t'$ and
      $(t,t')\in R_\mu$. 
      Observe that the disjunctive
      closure $\mu$ is  
      forward complete for $\pre$ iff 
      for any $s,t\in\Sigma$, if $s\in\pre(\mu(\{t\}))$ then 
      $\mu(\{s\})\subseteq\pre(\mu(\{t\}))$, and this happens iff
      for any $s,t\in\Sigma$, if $s\in \pre(\{t\})$ then 
      $\mu(\{s\})\subseteq\pre(\mu(\{t\}))$. This latter statement is
      equivalent to the fact that 
      for any 
      $s,s',t\in\Sigma$ such that $s\ra t$ and $s'\in\mu(\{s\})$, there exists
      $t'\in\mu(\{t\})$ such that $s'\ra t'$, namely,
      for any 
      $s,s',t\in\Sigma$ such that $s\ra t$ and $(s,s')\in R_\mu$, there exists
      $t'\in\Sigma$ such that $(t,t')\in R_\mu$ and $s'\ra t'$.
      
    \item[{\rm (2)}] $\mu\sqsubseteq\mu_\ell$ iff $R_\mu$ satisfies
      the property that for any $s,s'\in\Sigma$, if $(s,s')\in R_\mu$
      then $\ell(s)=\ell(s')$: In fact, 
      $\mu\sqsubseteq\mu_\ell$ $\:\Lra\:$ 
      $\forall s \in \Sigma.\: \mu(\{s\}) \subseteq
      \mu_\ell(\{s\})=[s]_\ell$ $\:\Lra\:$ 
      $\forall s,s' \in \Sigma.\: (s'\in \mu(\{s\}) \text{~implies~}
      s'\in [s]_\ell)$ $\:\Lra\:$ $\forall s,s'\in\Sigma.\: ((s,s')\in
      R_\mu \text{~implies~}
      \ell(s)=\ell(s'))$.
     
    \item[{\rm (3)}] Clearly, given $\mu'\in \ucod(\wp(\Sigma))$, 
$\mu\sqsubseteq\mu'$ iff $R_\mu\subseteq R_{\mu'}$.
      
  \end{itemize}

\noindent
  Let us show that $R_{\mu_\cK}=R_{\mathrm{sim}}$. By definition,
  $\mu_\cK$ is the most abstract disjunctive closure that is forward complete
  for $\pre$ and refines $\mu_\ell$. Thus, by the above points (1)
  and (2),
  it turns out that $R_{\mu_\cK}$ is a simulation on $\cK$. Consider 
  now 
  any  simulation $S$ on $\cK$ and the function
  $\mu'\ud \ok{\post_{S^*}}:\wp(\Sigma) \ra \wp(\Sigma)$. Let us notice that $\mu'\in
  \ucod(\wp(\Sigma))$ and
  $S\subseteq S^* = \ok{R_{\mu'}}$. Also, the relation $S^*$ is a simulation because
  $S$ is a simulation.
  Since $S^*$ is a simulation, we have that 
  $R_{\mu'}$ satisfies the conditions of the above points (1) and (2)
  so that $\mu'$ is forward complete for $\pre$ and $\mu'\sqsubseteq\mu_\ell$.
  Moreover, $\mu'$ is disjunctive so that  $\mu'$ is also forward complete for $\cup$.
  Thus, $\mu' \sqsubseteq \cS_{\cup,\pre}(\mu_\ell) =
  \mu_\cK$. Hence, by point (3) above, $R_{\mu'} \subseteq
  R_{\mu_\cK}$ so that $S\subseteq R_{\mu_\cK}$. We have therefore
  shown that $R_{\mu_\cK}$ is the largest simulation on $\cK$. 

  \noindent
  The fact that $P_{\mathrm{sim}} = \pr(\mu_\cK)$ 
  comes as a direct consequence because
  for any $s,t\in\Sigma$, $s\sim_{\mathrm{sim}} t$
  iff   $(s,t)\in R_{\mathrm{sim}}$ and $(t,s)\in R_{\mathrm{sim}}$.
  {}From $R_{\mu_\cK}=R_{\mathrm{sim}}$ we obtain that
  $s\sim_{\mathrm{sim}} t$ iff $s\in\mu_\cK(\{t\})$ and 
  $t\in\mu_\cK(\{s\})$ iff $\mu_\cK(\{s\})=\mu_\cK(\{t\})$. 
This holds iff $s$ and $t$ belong to the 
  same block in $\pr(\mu_\cK)$.
\end{proof}

Thus, the simulation preorder is characterized as the
forward complete shell of an initial abstract domain $\mu_\ell$
induced by the labeling $\ell$ w.r.t.\ set union $\cup$ and the
predecessor transformer $\pre$ while simulation equivalence is the
partition induced by this forward complete shell. 
Let us observe that set union and the predecessor
$\pre$ provide the semantics of, respectively, logical disjunction
and the existential next operator $\mathrm{EX}$. As shown in 
\cite{rt07}, simulation equivalence can be also
characterized in a precise meaning as the most abstract domain that
strongly preserves the language $$\varphi::= \mathit{atom}~|~\varphi_1
\wedge \varphi_2 ~|~ \varphi_1 \vee \varphi_2 ~|~ \mathrm{EX}
\varphi.$$

\begin{example}\label{esempio}
\rm
Let us consider the Kripke structure $\cK$
depicted below where the atoms $p$ and $q$ determine the
labeling function $\ell$.

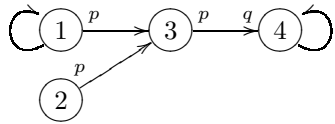
\begin{center}
\mbox{\xymatrix@R=8pt{
      *++[o][F]{1} \ar@(dl,ul)[] \ar[r]^(0.3){p}  &
      *++[o][F]{3}  \ar[r]^(0.3){p}^(0.7){q} & *++[o][F]{4}  \ar@(dr,ur)[] \\   
      *++[o][F]{2}\ar[ru] ^(0.25){p} & &
    }
  }
\end{center}
It is simple to observe that $P_{\mathrm{sim}} =
\{1,2,3,4\}$ because: (i)~while
$3\sra 4$ we have that
$1,2\not\in \pre (4)$ so that $1$ and $2$ are not simulation equivalent to $3$;
(ii)~while $1\sra 1$ we have that
$2\not\in \pre (12)$  so that $1$ is not simulation equivalent to $2$.

\noindent
The abstract domain induced by the
labeling is $\mu_\ell =\{\varnothing, 4,123,1234\}\in
\uco(\wp(\Sigma))$. As observed above, the forward complete shell
$\cS_{\cup,\pre}(\mu_\ell)=\cS_\cup (\cS_{\pre} (\mu_\ell))$ so that 
this domain can be obtained by iteratively closing  the image of
$\mu_\ell$
under $\pre$ and then by taking the
disjunctive completion:
\begin{itemize}
\item[--] $\mu_0 = \mu_\ell$;
\vspace*{-2pt}
\item[--] $\mu_1 = \Clc(\mu_0 \cup \pre(\mu_0))= \Clc(\mu_0 \cup
  \{\pre(\varnothing)=\varnothing,\, \pre(4)=34,\,
  \pre(123) = 12,\,$ $\pre(1234)=1234\}) = \{\varnothing,
  3,4, 12, 34, 123, 1234\}$;
\vspace*{-2pt}
\item[--] $\mu_2 = \Clc(\mu_1 \cup \pre(\mu_1))= \Clc(\mu_1 \cup
  \{\pre(3) = 12,\, \pre(12)=1,\,
   \pre(34)=1234\})=\{\varnothing,1,3,4, 12,34,123,
  1234\}$;
\vspace*{-2pt}
\item[--] $\mu_3 =  \Clc(\mu_2 \cup \pre(\mu_2))=\mu_2$ ~~~(fixpoint).
\end{itemize}
$\cS_{\cup,\pre}(\mu_\ell)$ is thus given by the disjunctive completion of
$\mu_2$, i.e., $\cS_{\cup,\pre}(\mu_\ell) = \{\varnothing,
1,3,4, 12,13,14,34,$ $123,124,134,1234\}=\mu_\cK$. Note that
$\mu_\cK(1)=1$, $\mu_\cK(2)=12$, $\mu_\cK(3)=3$ and $\mu_\cK(4)=4$. Hence, by Theorem~\ref{fcs},
the simulation preorder is 
$R_{\mathrm{sim}} = \{(1,1), (2,2), (2,1), (3,3), (4,4)\}$, while
$P_{\mathrm{sim}}=
\pr(\cS_{\cup,\pre}(\mu_\ell)) = \{1,2,3,4\}$. 
\qed 
\end{example}

Theorem~\ref{fcs} is one key result for proving
the correctness of our simulation algorithm
$\SA$ while it is not needed for understanding how
$\SA$ works and how to implement it efficiently. 

\section{Partition-Relation Pairs}\label{prp}
Let $P\in \Part(\Sigma)$ and $R\subseteq P\times P$
be any relation on the partition $P$. One such pair $\tuple{P,R}$ is called a
\emph{partition-relation pair}.  
A partition-relation pair $\tuple{P,R}$ induces a disjunctive closure
$\mu_{\tuple{P,R}}\in \ucod(\wp(\Sigma)_\subseteq)$ as follows: for any $X\in
\wp(\Sigma)$,
$$\mu_{\tuple{P,R}} (X) \ud
\cup\{ C\!\in\! P\, |\, \exists B\!\in\! P.\, B\cap X \!\neq\! \varnothing, (B,C)\in
R^*\}.$$ 
It is easily shown that $\mu_{\tuple{P,R}}$ is indeed
a disjunctive uco.  Note that, for any $B\in P$ and $x\in B$, 
$$\mu_{\tuple{P,R}} (\{x\}) =
\mu_{\tuple{P,R}} (B) = \cup R^*(B)=\cup \{ C\in P~|~ (B,C)\in
R^*\}.$$
  
This correspondence is a key  logical point for proving the correctness 
of our simulation
algorithm. In fact, our algorithm maintains a partition-relation pair,
where the relation is a preorder, and our proof of correctness
depends on the fact that this partition-relation pair logically represents a
corresponding disjunctive abstract domain.

\begin{example}\label{simplebis}
Let $\Sigma =\{1,2,3,4\}$, 
$P= \{12,3,4\}\in \Part(\Sigma)$ 
and $R=\{(12,3), (3,4), (4,3)\}$. Note that
$R^*=\{(12,12), (12,3), (12,4), (3,3), (3,4), (4,3), (4,4)\}$. 
The disjunctive abstract domain $\mu_{\tuple{P,R}}$ 
is such that $\mu_{\tuple{P,R}}(\{1\}) = \mu_{\tuple{P,R}}(\{2\}) 
=\{1,2,3,4\}$ and $\mu_{\tuple{P,R}} (\{3\})= \mu_{\tuple{P,R}} (\{4\})
=\{3,4\}$, so that the 
image of 
$\mu_{\tuple{P,R}}$ is $\{\varnothing, 34, 1234\}$.   
 \qed
\end{example}

On the other hand, any abstract domain $\mu \in
\uco(\wp(\Sigma))$ induces a partition-relation pair $\tuple{P_\mu,
  R_\mu}$ as follows: 
\begin{itemize}
\item[--] $P_\mu \ud \pr(\mu)$;
\item[--] $R_\mu \ud \{(B,C)\in P_\mu\times P_\mu~|~C \subseteq \mu(B)\}$. 
\end{itemize}

The following properties of partition-relation pairs will be useful in
later proofs. 
\begin{lemma}\label{prprop}
Let $\tuple{P,R}$ be a partition-relation pair and $\mu \in
\uco(\wp(\Sigma))$. 
\begin{itemize}
\item[{\rm (i)}] $P\preceq
\pr(\mu_{\tuple{P,R}})$. 

\item[{\rm (ii)}]
$\tuple{P_\mu,R_\mu} = \tuple{P_{\mud},R_{\mud}}$.
\end{itemize}
\end{lemma}
\begin{proof}
(i)~~We already observed above that if $B\in P$ and $x\in B$ then 
$\mu_{\tuple{P,R}} (\{x\})  =
\mu_{\tuple{P,R}} (B)$, so that $B\subseteq \{y\in
\Sigma~|~\mu_{\tuple{P,R}} (\{x\})  = \mu_{\tuple{P,R}} (\{y\})\}$
which is a block in  $\pr(\mu_{\tuple{P,R}})$. \\
(ii)~~By Lemma~\ref{prop}~(iv),
$P_\mu=\pr(\mu)=\pr(\mud)=P_{\mud}$. Moreover, 
\begin{align*}
R_\mu &=\text{~~~~~[by definition]}\\
\{(B,C)\in P_\mu \times P_\mu ~|~ C\subseteq \mu(B)\} & =
\text{~~~~~[as $P_\mu = P_\mud$]}\\ 
\{(B,C)\in P_\mud \times P_\mud ~|~ C\subseteq \mu(B)\} & =
\text{~~~~~[as $\mu(B)=\mud(B)$]} \\
\{(B,C)\in P_\mud \times P_\mud ~|~ C\subseteq \mud(B)\} & =
\text{~~~~~[by definition]}\\
R_\mud &. \qedhere
\end{align*}
\end{proof}

It turns out that 
the above two correspondences between partition-relation pairs and
disjunctive abstract domains are inverse of each other when the relation is
a partial order. 

\begin{lemma}\label{luno}
For any partition $P\in \Part(\Sigma)$, partial order
$R\subseteq P\times P$ and disjunctive
abstract domain $\mu\in \ok{\ucod(\wp(\Sigma))}$, we have that
$\ok{\tuple{P_{\mu_{\tuple{P,R}}},R_{\mu_{\tuple{P,R}}}}} = \tuple{P,R}$ and
$\ok{\mu_{\tuple{P_\mu,R_\mu}}} = \mu$. 
\end{lemma}
\begin{proof}
Let us show that
$\tuple{P_{\mu_{\tuple{P,R}}},R_{\mu_{\tuple{P,R}}}} = \tuple{P,R}$.
We first prove that $P_{\mu_{\tuple{P,R}}}=P$, i.e.\ 
$\pr(\mu_{\tuple{P,R}})=P$. On the one hand, 
by Lemma~\ref{prprop}~(i), 
$P\preceq \pr(\mu_{\tuple{P,R}})$. On the other hand, if $x,y\in
\Sigma$,  
$\mu_{\tuple{P,R}}(\{x\}) = \mu_{\tuple{P,R}}(\{y\})$ and $x\in B_x\in
P$ and $y\in B_y\in P$ then $(B_x,B_y)\in R^*$ and $(B_y,B_x)\in
R^*$. Since $R$ is a partial order, we have that $R^*=R$ is a partial order
as well, so that $B_x = B_y$, namely
$\pr(\mu_{\tuple{P,R}})\preceq P$. 

\noindent
Let us prove now that $\ok{R_{\mu_{\tuple{P,R}}}}=R$. In fact, for any
$(B,C) \in \pr(\mu_{\tuple{P,R}})\times \pr(\mu_{\tuple{P,R}})$, 
\begin{align*}
(B,C)\in \ok{R_{\mu_{\tuple{P,R}}}} & \Lra \text{~~~~[by definition of 
$\ok{R_{\mu_{\tuple{P,R}}}}$]}\\
C\subseteq \ok{\mu_{\tuple{P,R}}(B)} &\Lra \text{~~~~[by definition of 
$\ok{\mu_{\tuple{P,R}}}$]}\\
(B,C)\in R^* &\Lra \text{~~~~[since $R^*=R$]}\\
(B,C)\in R.&
\end{align*}

\noindent
Finally, let us show that $\mu_{\tuple{P_\mu,R_\mu}} = \mu$. Since 
both $\mu_{\tuple{P_\mu,R_\mu}}$ and $\mu$ are disjunctive it is
enough to prove that for all $x\in \Sigma$, 
$\mu_{\tuple{P_\mu,R_\mu}}(\{x\}) = \mu(\{x\})$. Given $x\in \Sigma$
consider the block $B_x\in P_\mu=\pr(\mu)$ containing $x$. Then, 
\begin{align*}
\mu_{\tuple{P_\mu,R_\mu}}(\{x\}) & = \text{~~~~[by definition of 
$\mu_{\tuple{P_\mu,R_\mu}}$]}\\
\cup \{C\in P_\mu~|~ (B_x,C)\in R_\mu^*\} & = \text{~~~~[since $R_\mu^*
  = R_\mu$]}\\
\cup \{C\in P_\mu~|~ (B_x,C)\in R_\mu\} & = \text{~~~~[by definition of 
$R_\mu$]}\\ 
\cup \{C\in P_\mu~|~ C \subseteq \mu(B_x)\} & = \text{~~~~[by Lemma~\ref{prop}~(ii)]}\\ 
\mu(B_x) & = \text{~~~~[since
  $\mu(B_x)=\mu(\{x\})$]}\\
\mu(\{x\})&. \qedhere
\end{align*}
\end{proof}

Our simulation algorithm relies on the following
condition on a partition-relation pair $\tuple{P,R}$ w.r.t.\ a
transition system $(\Sigma,\sra)$ which guarantees
that the corresponding disjunctive abstract domain $\mu_{\tuple{P,R}}$
is forward complete for the predecessor $\pre$.

\begin{lemma}\label{ltre}
Let $(\Sigma,\sra)$ be a transition system and
$\tuple{P,R}$ be a partition-relation pair where $R$ is reflexive.
Assume that for any $B,C\in P$, if 
$C\cap\pre(B)\not=\varnothing$ then
$\cup R(C)\subseteq\pre(\cup R(B))$.
Then, $\mu_{\tuple{P,R}}$ is  forward
complete for $\pre$.
\end{lemma}
\begin{proof}
We preliminarily show the following fact: 

\begin{itemize}
\item[$(\ddagger)$] 
Let $\mu\in \ucod(\wp(\Sigma))$ and $P\in \Part(\Sigma)$ such that $P\preceq \pr(\mu)$. 
Then, $\mu$ is forward
  complete for $\pre$ iff for any $B,C\in P$, if 
  $C\cap\pre(B)\not=\varnothing$ then
  $\mu(C)\subseteq\pre(\mu(B))$. 

\smallskip
\noindent
($\Rightarrow$) Let $B,C \in P$ 
  such that $C\cap\pre(B)\not=\varnothing$. Since $B\subseteq \mu(B)$ we also have that
  $C\cap\pre(\mu(B))\not=\varnothing$. 
By forward completeness, $\pre(\mu(B)) = \mu(\pre(\mu(B))$.
Since $P\preceq
\pr(\mu)$, $C\in P$ and $C\cap \mu(\pre(\mu(B))) = C \cap \pre(\mu(B))\neq \varnothing$
we have that $C\subseteq \mu(\pre(\mu(B))) = \pre(\mu(B))$, so that, by applying 
the monotone map $\mu$, $\mu(C) \subseteq \mu(\pre(\mu(B))) = \pre(\mu(B))$.

\noindent
($\Leftarrow$)
  Firstly, we show the following property $(*)$: for any $B,C\in P$,
  if $C\cap\pre(\mu(B))\not=\varnothing$ then 
  $\mu(C)\subseteq\pre(\mu(B))$. 
  Since $P\preceq \pr(\mu)$, by Lemma~\ref{prop}~(ii),
$C\cap\pre(\mu(B)) = C\cap \pre(\cup \{D \in P~|~ D
\subseteq \mu(B)\})$, so that 
if $C\cap\pre(\mu(B))\neq \varnothing$ then
  $C\cap\pre(D)\not=\varnothing$ for some $D\in P$ such that $D\subseteq \mu(B)$. 
Hence, by hypothesis, 
  $\mu(C)\subseteq\pre(\mu(D))$. Since $\mu(D)\subseteq \mu(B)$, 
we thus obtain  that $\mu(C) \subseteq \pre(\mu(D)) 
\subseteq \pre(\mu(B))$. 
Let us now prove that $\mu$ is forward complete for $\pre$.  We first show the
following property $(**)$: for
any $B\in P$, $\mu(\pre(\mu(B)))\subseteq\pre(\mu(B))$. In fact,
since
$P\preceq \pr(\mu)$, we have that:
\begin{align*} 
\mu(\pre(\mu(B))) & = \text{~~~~~[by Lemma~\ref{prop}~(iii) because $\mu$
  is additive]}\\
\cup \{ \mu(C) ~|~C\in P,\: C \cap \pre(\mu(B))\neq \varnothing\} 
 & \subseteq \text{~~~~~[by the above property $(*)$]}\\
\pre(\mu(B)). & 
\end{align*} 
Hence, for any $X\in \wp(\Sigma)$, we have that:
\begin{align*} 
\mu(\pre(\mu(X))) & =  \text{~~~~~[since, by Lemma~\ref{prop}~(iii), 
$\mu(X) = \cup_i \mu(B_i)$ for some $\{B_i\}\subseteq P$]}\\
\mu(\pre(\cup_i \mu(B_i))) & = \text{~~~~~[since $\mu$ and $\pre$ are additive]}\\
\cup_i \mu(\pre(\mu(B_i))) & \subseteq \text{~~~~~[by the above property $(**)$]}\\
\cup_i \pre(\mu(B_i)) & = \text{~~~~~[since $\pre$ is additive]}\\
\pre(\cup_i \mu(B_i)) & = \text{~~~~~[since $\mu(X) = \cup_i \mu(B_i)$]}\\
\pre(\mu(X))&. 
\end{align*} 
\end{itemize}

\noindent
Let us now turn to show the lemma.
By Lemma~\ref{prprop}~(i), we have that $P\preceq \pr(\mu_{\tuple{P,R}})$. 
By the above fact~$(\ddagger)$, in order to prove that 
$\mu_{\tuple{P,R}}$ is forward complete for $\pre$ it is sufficient to show that
for any $B,C\in P$, 
if  $C\cap\pre(B)\not=\varnothing$ then
  $\mu_{\tuple{P,R}}(C)\subseteq\pre(\mu_{\tuple{P,R}}(B))$. 
Thus, assume that 
  $C\cap\pre(B)\not=\varnothing$. We need to show that 
$\cup R^*(C) \subseteq \pre(\cup R^* (B))$. 
Assume that $(C,D)\in R^*$, namely that there exist $\{B_i\}_{i\in [0,k]}\subseteq P$, 
for some $k\geq 0$, such that $B_0 = C$, $B_k=D$ and for any $i\in [0,k)$,
$(B_i,B_{i+1})\in R$. We show by induction on $k$ that $D \subseteq 
\pre(\cup R^* (B))$. 
\begin{description}
\item[($k=0$)] This means that $C=D$. Since $R$ is assumed to be
  reflexive, we have that $(C,C)\in R$. By hypothesis, $\cup R(C) \subseteq
\pre(\cup R(B))$ so that
we obtain $D=C\subseteq \cup R(C) \subseteq
\pre(\cup R(B))\subseteq \pre(\cup R^*(B))$.  
\item[($k+1$)] Assume that $(C,B_1),(B_1,B_2),...,(B_k,D) \in R$. By inductive hypothesis,
$B_k\subseteq \pre(\cup R^* (B))$. Note that, by additivity of $\pre$, 
$\pre(\cup R^*(B)) = \cup \{\pre(E)~|~ 
E\in P,\: (B,E)\in R^*\}$.  
Thus, there exists some $E\in P$ such that $(B,E) \in
R^*$ and $B_k \cap \pre(E)\neq \varnothing$. Hence, by hypothesis,
$\cup R(B_k) \subseteq
\pre(\cup R(E))$. Observe that $\cup R(E)\subseteq \cup R^*(E)
\subseteq \cup R^*
(B)$ so that $D\subseteq \cup R(B_k) \subseteq \pre(\cup R(E))
\subseteq \pre(\cup R^*
(B))$. \qedhere 
\end{description}
\end{proof}

\afterpage{\clearpage}

\begin{algorithm}[t]
\small
\SetVline
\Setnlskip{1.5em}
\SetAlTitleFnt{textsc}
\Indm
$\mathit{SchematicSimilarity}()$ $\{$

\Indp
\lForAll{$v\in \Sigma$}{$\Sim(v):= [v]_\ell$\;}
\While{$\exists u,v,w \in \Sigma$ \KwSty{such that} $(u\sra v \;\, \&\;\, w\!\in\! \mathit{Sim}(u)\;\,\&\;\,
\posts(\{w\})\cap \mathit{Sim(v)}=\varnothing)$}{
$\mathit{Sim}(u):=\mathit{Sim}(u)\smallsetminus \{w\}$\;
}
\Indm
$\}$

\BlankLine
\BlankLine
\BlankLine
$\mathit{RefinedSimilarity}()$ $\{$

\Indp
\ForAll{$v\in \Sigma$}{
$\mathit{prevSim}(v):=\Sigma$\;
\lIf{$\post(\{v\})=\varnothing$}{$\mathit{Sim}(v):= [v]_\ell$;
  ~\KwSty{else}~ $\mathit{Sim}(v):= [v]_\ell\cap \pre(\Sigma)$\;}
}

\While{$\exists v \in \Sigma$ \KwSty{such that} $\mathit{Sim}(v) \neq \mathit{prevSim(v)})$}{
\tcp*[h]{\textrm{~$\Inv_1$:} $\forall v \in \Sigma.\;\mathit{Sim}(v)\subseteq 
\mathit{prevSim}(v)$}

\tcp*[h]{\textrm{~$\Inv_2$:} $\forall u,v\in \Sigma.\;
u\sra v \,\Rightarrow \mathit{Sim}(u)\subseteq\pre(\mathit{prevSim}(v))$}

$\mathit{Remove}:= \pre(\mathit{prevSim}(v))\smallsetminus \pre(\mathit{Sim}(v))$\;
$\mathit{prevSim}(v):= \mathit{Sim}(v)$\;
\lForAll{$u \in \pre(v)$}{$\mathit{Sim}(u):=\mathit{Sim}(u)\smallsetminus \mathit{Remove}$}\;
}
\Indm
$\}$

\BlankLine \BlankLine \BlankLine
$\HHK()$ $\{$

\Indp
\tcp*[h]{\lForAll{$v \in \Sigma$}{$\mathit{prevSim}(v):=\Sigma;$}}

\ForAll{$v\in \Sigma$}{
\lIf{$\post(\{v\})=\varnothing$}{$\mathit{Sim}(v):= [v]_\ell$;
  ~\KwSty{else}~ $\mathit{Sim}(v):= [v]_\ell\cap \pre(\Sigma)$\;}
$\mathit{Remove}(v):= \pre(\Sigma) \smallsetminus \pre(\mathit{Sim}(v))$\;
}

\While{$\exists v \in \Sigma$ \KwSty{such that} $\mathit{Remove}(v) \neq \varnothing$}{
\tcp*[h]{\textrm{~$\Inv_3$:} $\forall v\in \Sigma.\; \mathit{Remove}(v)=\pre(\mathit{prevSim}(v))\smallsetminus \pre(\mathit{Sim}(v))$}

\tcp*[h]{$\;\mathit{prevSim}(v) := \mathit{Sim}(v);$}

$\mathit{Remove}:=\mathit{Remove}(v)$\;
$\mathit{Remove}(v):= \varnothing$\;
\ForAll{$u \in \pre(v)$}{
	\ForAll{$w \in \mathit{Remove}$}{
	\If{$w\in \mathit{Sim}(u)$} {
		$\mathit{Sim}(u):=\mathit{Sim}(u)\smallsetminus\{w\}$\;
		\ForAll{$w''\in \pre(w)$ \KwSty{such that} $w''\not\in \pre(\mathit{Sim}(u)$ }
			{$\mathit{Remove}(u):=\mathit{Remove}(u)\cup\{w''\};$}
	}
}
}
}
\Indm
$\}$
\caption{$\HHK$ Algorithm.}\label{hhkfig}
\end{algorithm}

\section{Henzinger, Henzinger and Kopke's Algorithm}

Our simulation algorithm $\SA$ is designed as a
symbolic modification of Henzinger, Henzinger and Kopke's
simulation algorithm \cite{hhk95}.  This 
algorithm is designed in three incremental steps encoded by 
the procedures
$\mathit{SchematicSimilarity}$, $\mathit{RefinedSimilarity}$ and
$\HHK$ (called $\mathit{EfficientSimilarity}$ in \cite{hhk95}) in
Figure~\ref{hhkfig}. 

Consider any (possibly non total) 
finite Kripke structure $(\Sigma,\sra, \ell)$. 
The idea of the basic $\mathit{SchematicSimilarity}$ 
algorithm is simple. 
For each state $v\in \Sigma$, the simulator set $\Sim(v)\subseteq \Sigma$ contains
states that are candidates for simulating $v$. Hence, $\Sim(v)$ is
initialized with all the states having the same labeling as
$v$, that is $[v]_\ell$. The algorithm then proceeds iteratively as follows: if
$u\sra v$,  $w\in \Sim(u)$ but there is no $w'\in \Sim(v)$ such that
$w\sra w'$ then $w$ cannot simulate $u$ and therefore $\Sim(u)$ is
refined to $\Sim(u)\smallsetminus \{w\}$. 

This basic procedure is then refined to the algorithm
$\mathit{RefinedSimilarity}$.  The key point
here is to store for each state $v\in \Sigma$ an additional set of states
$\prevSim(v)$ that is a superset of $\Sim(v)$ (invariant $\Inv_1$) and
contains the states that were in $\Sim(v)$ in some past iteration where
$v$ was selected. If
$u\sra v$ then the invariant $\Inv_2$ allows to refine $\Sim(u)$ by
scrutinizing only the states in $\pre(\prevSim(v))$ instead of all
the possible states in $\Sigma$: 
In fact, while in $\mathit{SchematicSimilarity}$, 
$\Sim(u)$ is reduced to  $\Sim(u)\smallsetminus (\Sigma \smallsetminus
\pre(\mathit{Sim}(v))$, in $\mathit{RefinedSimilarity}$,
$\Sim(u)$ is reduced in the same way by removing from it the states
in $\mathit{Remove} \ud \pre(\mathit{prevSim}(v))\smallsetminus
\pre(\mathit{Sim}(v))$. The initialization of $\Sim(v)$ that
distinguishes the case $\post(\{v\})=\varnothing$ allows to  initially establish 
the invariant  $\Inv_2$. 
Let us remark that the
original $\mathit{RefinedSimilarity}$ algorithm presented in~\cite{hhk95}
contains the following bug: the statement $\prevSim(v):=\Sim(v)$ is
placed just after the inner for-loop instead of immediately preceding the
inner for-loop. It turns out that this 
is not correct as shown by the following
example.

\begin{example}\label{contro}
\rm
Let us consider the Kripke structure in Example~\ref{esempio}. We
already observed that the simulation relation is $R_{\mathrm{sim}} =
\{(1,1), (2,2), (2,1), (3,3), (4,4)\}$. 
However, one can check that
the original version 
of the $\mathit{RefinedSimilarity}$ algorithm  in \cite{hhk95}~---~where the assignment
$\prevSim(v):=\Sim(v)$ follows the inner for-loop~---~provides as output
$\Sim(1)=\{1,2\}$, $\Sim(2)=\{1,2\}$, $\Sim(3)=\{3\}$,
$\Sim(4)=\{4\}$, namely the state $2$ appears to simulate the state
$1$ while this is not the case. 
The problem with the original version
in \cite{hhk95} of the 
$\mathit{RefinedSimilarity}$ algorithm lies in the fact that when
$v\in \pre(\{v\})$~---~like in this example for state 1~---~it may
happen that during the inner for-loop the set $\Sim(v)$ is
refined to $\Sim(v)\smallsetminus \Remove$ so that if 
the assignment $\prevSim(v):=\Sim(v)$ follows the inner for-loop then $\prevSim(v)$ might be
computed as an incorrect subset of the right set. 
\qed 
\end{example}

$\mathit{RefinedSimilarity}$ is further refined to the final
$\HHK$ algorithm.  The idea here is that
instead of recomputing at each iteration of the while-loop the set
$\mathit{Remove} := \pre(\mathit{prevSim}(v))\smallsetminus
\pre(\mathit{Sim}(v))$ for the selected state $v$, a set $\Remove(v)$
is maintained and incrementally updated for each state $v\in \Sigma$ in such a
way that it satisfies the invariant $\Inv_3$. The original version of
$\HHK$ in \cite{hhk95} also suffers from a bug
that is a direct consequence of the problem in
$\mathit{RefinedSimilarity}$ described above: within the main
while-loop of $\HHK$, the statement $\Remove(v):=\varnothing$
is placed just after the outermost for-loop instead of immediately preceding the
outermost for-loop. It is easy to show that this is not correct by resorting
again to Example~\ref{contro}. 

The implementation of $\HHK$
exploits a matrix $\mathit{Count}(u,v)$, indexed on states $u,v\in
\Sigma$, such that $\mathit{Count}(u,v)=|\post (u) \cap \Sim(v)|$,
i.e., $\mathit{Count}(u,v)$ stores the number of transitions from $u$
to some state $w\in \Sim(v)$. 
Hence, the test $w'' \not \in \pre(\mathit{Sim}(u))$ in the
innermost for-loop can be done in $O(1)$ by checking whether
$\mathit{Count}(w'',u)$ is 0 or not.  This provides an efficient
implementation of $\HHK$ that runs in
$O(|\Sigma||\sra|)$ time, while the space complexity is in
$O(|\Sigma|^2 \log |\Sigma|)$, namely it is more than
quadratic in the size of the state space. Let us remark that the key
property for showing the $O(|\Sigma||\sra|)$ time bound is as follows:
if a state $v$ is selected at some iterations $i$ and $j$ of the
while-loop and the iteration $i$ precedes the iteration $j$ 
then $\mathit{Remove}_i(v) \cap
\mathit{Remove}_j (v)=\varnothing$, so that the sets in
$\{\Remove_i(v)~|~ v$ is selected at some iteration i $\}$ are
pairwise disjoints.

\section{A New Simulation Algorithm}

\subsection{The Basic Algorithm}
Let us consider any (possibly non total) 
finite Kripke structure $(\Sigma,\sra, \ell)$.
As recalled above, the $\HHK$ procedure maintains for each state
$s\in \Sigma$ a simulator set $\Sim(s)\subseteq \Sigma$ and a remove
set $\Remove(s)\subseteq \Sigma$. The simulation preorder 
$R_{\mathrm{sim}}$ is encoded by the output
$\{\Sim(s)\}_{s\in \Sigma}$ as follows: $(s,s')\in R_{\mathrm{sim}}$
iff $s'\in \Sim(s)$. Hence, 
the simulation equivalence
partition $P_{\mathrm{sim}}$ is obtained as follows: $s$ and $s'$ are
simulation equivalent iff $s\in \Sim(s')$ and $s'\in \Sim(s)$.  
Our algorithm relies on the idea of modifying
the $\HHK$ procedure in order to maintain a partition-relation pair
$\tuple{P,\Rel}$ in place of  $\{\Sim(s)\}_{s\in \Sigma}$,
together with a remove set $\Remove(B)\subseteq \Sigma$ for each block $B\in P$. The
basic idea is to replace the family of sets $\cS=\{\Sim(s)\}_{s\in \Sigma}$
with the following state partition $P$ induced by
$\cS$: $s_1 \sim_\cS s_2$ iff for all $s\in \Sigma$, $s_1 \in
\Sim(s)\Lra s_2 \in \Sim(s)$.  
Then, a reflexive relation $\Rel\subseteq P \times
P$ on $P$ gives rise to a partition-relation pair where the
intuition is as follows:
given a state $s$ and a block $B\in P$
(i)~if $s\in B$ then 
the current simulator set for $s$ is a the union of blocks in $P$ that 
are in relation with $B$, i.e.\
$\Sim(s) = \cup\! \Rel(B)$;  (ii) if $s,s'\in B$ then 
$s$ and $s'$ are currently candidates to be simulation equivalent. 
Thus, a partition-relation pair
$\tuple{P,\Rel}$ represents the current
approximation of the simulation preorder and in particular $P$ represents the
current approximation of simulation
equivalence.  

Partition-relation pairs have been used by Henzinger, Henzinger and
Kopke's~\cite{hhk95} to compute the simulation preorder on effectively
presented infinite transition systems, notably hybrid automata.
Henzinger et al.\ provide a symbolic procedure, called
$\mathit{SymbolicSimilarity}$ in \cite{hhk95}, that is derived as a symbolization through
partition-relation pairs of their basic simulation algorithm
$\mathit{SchematicSimilarity}$ in Figure~\ref{hhkfig}.
Moreover,
partition-relation pairs  are 
also exploited by
Gentilini et al.~\cite{gpp03} in their simulation
algorithm for representing simulation relations. 
The distinctive feature of our use of partition-relation pairs is that,
by relying on the results in Section~\ref{prp}, we logically view
partition-relation pairs 
as abstract domains and therefore we can reason on them by using
abstract interpretation.

Following Henzinger et al.\ \cite{hhk95}, our simulation
algorithm is designed in three incremental steps. We exploit the following
results for designing the basic algorithm. 

\begin{itemize}
\item[--] Theorem~\ref{fcs} tells us that the simulation
preorder can be obtained from the forward $\{\cup,\pre\}$-complete shell of
an initial abstract domain $\mu_\ell$ induced by the labeling $\ell$. 

\item[--]
As shown in
Section~\ref{prp}, a partition-relation pair can be viewed
as representing a disjunctive abstract domain. 
\item[--] Lemma~\ref{ltre}
gives us a condition on a partition-relation pair which guarantees
that the corresponding abstract domain is forward complete for
$\pre$. Moreover, this abstract domain is disjunctive as well, being
induced by a partition-relation pair.
\end{itemize}

\incmargin{1em}
\linesnumbered
\begin{algorithm}[t]
\small
\SetVline
\Setnlskip{1.5em}
\SetAlTitleFnt{textsc}
\Indm
$\BasicSA(\mathit{PartitionRelation} ~ \tuple{P,\mathit{Rel}})$ $\{$

\Indp
\While{$\exists B,C \in P$ \KwSty{such that} $(C\cap\pre(B)\neq\varnothing\;\,\&\;\cup\!\mathit{Rel}(C)\not\subseteq\pre(\cup\mathit{Rel}(B)))$} {
$S:=\pre(\cup\mathit{Rel}(B))$\;
$P_{\mathrm{prev}} := P$; ~$B_{\mathrm{prev}} :=B$\;
$P := \mathit{Split}(P,S)$\;
\lForAll{$C \in P$}
{$Rel(C):=\{D\in P~|~D\subseteq\cup \mathit{Rel}(\parent_{P_{\mathrm{prev}}}(C))\}$;}

\lForAll{$C\in P$ \KwSty{such that} $C\cap\pre(B_{\mathrm{prev}})\neq\varnothing$}
{$Rel(C):=\{D\in Rel(C) ~|~ D\subseteq S\}$;}
}
\Indm
$\}$
\caption{Basic Simulation Algorithm.}\label{fig:basic}
\end{algorithm}

Thus, the idea
consists in
iteratively and minimally refining an initial
partition-relation pair $\tuple{P,\Rel}$ induced by the labeling
$\ell$ until the condition of Lemma~\ref{ltre} is
satisfied: for all $B,C\in P$,
$$C\cap \pre(B)\neq \varnothing\:\Ra\:
\cup\!\Rel(C) \subseteq \pre(\cup\!\Rel(B)).$$
Let us observe that $C\cap \pre(B)\neq \varnothing$ means that
$C\sra^{\exists\exists} B$.
The basic algorithm, called
$\BasicSA$, is in Figure~\ref{fig:basic}.  The
current partition-relation pair $\tuple{P,\Rel}$ is refined by the
following three steps in $\BasicSA$. If $B$ is the block of
the current partition $P$ selected by the while-loop then:

\begin{itemize}
\item[{\rm (i)}]
the current partition $P$ is split with respect to the set
$S=\pre(\cup\!\Rel(B))$;
\item[{\rm (ii)}] if $C$ is a newly generated block after splitting
  the current partition 
  and $\parent_{P_{\mathrm{prev}}}(C)$ is its parent block in the
  partition $P_{\mathrm{prev}}$ before the splitting
  operation then $\Rel(C)$ is modified so as that $\cup \!\Rel(C) =
  \cup\!\Rel(\parent_{P_{\mathrm{prev}}}(C))$;
\item[{\rm (iii)}] the current relation $\Rel$ is refined for the (new and old)
  blocks $C$ such that $C\sra^{\exists\exists} B$ by removing from
  $\Rel(C)$ those blocks that are not contained in $S$;
  observe that after having split $P$ w.r.t.\
  $S$ it turns out that one such block $D$ either is contained
  in $S$ or is disjoint with $S$.
\end{itemize}

Let us remark that although the symbolic simulation algorithm
for infinite graphs  
$\mathit{SymbolicSimilarity}$ in \cite{hhk95} 
may appear similar to our $\BasicSA$ algorithm, it is instead   
inherently different due to the following reason:
the role played by the condition:
$C\sra^{\exists\exists} B \;\&\:
\cup\!\Rel(C) \not\subseteq \pre(\cup\!\Rel(B))$
in the while-loop of $\BasicSA$ is played in $\mathit{SymbolicSimilarity}$
by: $C\sra^{\exists\exists} \cup\!\Rel(B)\;\&\:
\cup\!\Rel(C) \not\subseteq \pre(\cup\!\Rel(B))$, and this latter
condition is computationally harder to check. 

The following correctness result formalizes that 
$\BasicSA$ can be viewed as an abstract domain refinement 
algorithm that allows us to compute forward complete
shells for $\{\cup,\pre\}$. 
For any abstract domain $\mu \in \uco
(\wp(\Sigma))$,  we write $\mu'=
\BasicSA(\mu)$ when the algorithm $\BasicSA$ on an
input partition-relation $\tuple{P_{\mu},R_\mu}$ terminates and outputs a
partition-relation pair $\tuple{P',R'}$ such that
$\mu'=\mu_{\tuple{P',R'}}$.

\begin{theorem}\label{mainAlgo}
Let $\Sigma$ be finite. Then, $\BasicSA$ terminates on any
input domain $\mu \in \uco(\wp(\Sigma))$ and $\BasicSA(\mu)=
\cS_{\cup,\pre}(\mu)$.
\end{theorem}
\begin{proof}
Let $\tuple{P_{\mathrm{curr}},R_{\mathrm{curr}}}$ and
$\tuple{P_{\mathrm{next}},R_{\mathrm{next}}}$ be, respectively, the
current and next partition-relation pair in some iteration of 
$\BasicSA(\mu)$. By line 5, $P_{\mathrm{next}}\preceq
P_{\mathrm{curr}}$ always holds. Moreover, if $P_{\mathrm{next}} =
P_{\mathrm{curr}}$ then it turns out that $R_{\mathrm{next}} \subsetneq
R_{\mathrm{curr}}$: in fact, if $B,C\in P_{\mathrm{curr}}$, $C\cap
\pre(B)\neq \varnothing$ and $\cup R_{\mathrm{curr}}(C)
\not\subseteq \pre(\cup R_{\mathrm{curr}}(B))$ then, by lines 6 and 7, 
$\cup R_{\mathrm{next}}(C)
\subsetneq \cup R_{\mathrm{curr}}(C)$ because there exists $x\in 
\cup R_{\mathrm{curr}}(C)$ such that $x\not\in\pre(\cup
R_{\mathrm{curr}}(B))$ so that if $B_x\in P_{\mathrm{next}} =
P_{\mathrm{curr}}$ is the block that contains $x$ then $B_x\cap ( 
\cup R_{\mathrm{next}}(C)) = \varnothing$ while $B_x \subseteq 
\cup R_{\mathrm{curr}}(C)$. Thus, either 
$P_{\mathrm{next}}\prec
P_{\mathrm{curr}}$ or $R_{\mathrm{next}} \subsetneq
R_{\mathrm{curr}}$, so that, since the state space $\Sigma$ is finite,
the procedure
$\BasicSA$ terminates. 

\noindent
Let $\mu'=  \BasicSA(\mu)$, namely, let
$\mu'=\mu_{\tuple{P',R'}}$ where $\tuple{P',R'}$ is the output of
$\BasicSA$ on input $\tuple{P_{\mu},R_\mu}$. 
Let $\{\tuple{P_i,R_i}\}_{i\in [0,k]}$ be the sequence of
partition-relation pairs computed by $\BasicSA$, where
$\tuple{P_0,R_0}= \tuple{P_\mu,R_\mu}$ and $\tuple{P_k,R_k}=\tuple{P',R'}$.  
Let us first observe that for any $i\in [0,k)$, $P_{i+1} \preceq P_i$
because the current partition is refined by the splitting operation in
line 5. Moreover, for any $i\in [0,k)$ and $C\in P_{i+1}$, note that
$\cup {R_{i+1}}(C) \subseteq \cup {R_i}(\parent_{P_i}(C))$, because the
current relation is modified only at lines 6 and 7. 

\noindent
Let us also observe that for
any $i\in [0,k]$, $R_i$ is a reflexive relation because
$R_0$ is reflexive  and
the operations at lines 6-7 preserve the reflexivity 
of the current
relation. Let us show this latter fact. 
If $C\in P_{\mathrm{next}}$ is such that $C\cap
\pre(B_{\mathrm{prev}}) \neq \varnothing$ then because, by 
hypothesis, $B_{\mathrm{prev}}\in
R_{\mathrm{prev}}(B_{\mathrm{prev}})$, we have that  $C\cap
\pre(\cup {R_{\mathrm{prev}}}(B_{\mathrm{prev}}))\neq \varnothing$ so that
$C\subseteq S=\pre(\cup
{R_{\mathrm{prev}}}(B_{\mathrm{prev}}))$. Hence, if $C\in 
P_{\mathrm{next}}\cap P_{\mathrm{prev}}$ then $C\in
R_{\mathrm{next}}(C)$, while  
if $C\in P_{\mathrm{next}}\smallsetminus P_{\mathrm{prev}}$
then, by hypothesis, $\parent_{P_{\mathrm{prev}}}(C) \in
R_{\mathrm{prev}}(\parent_{P_{\mathrm{prev}}}(C))$ so that, by line 6, 
$C\in R_{\mathrm{next}}(C)$ also in this case.

\noindent
For any $B\in P'=P_k$, we have that 
\begin{align*}
\mu'(B) &=\text{~~~[by definition of $\mu'$]}\\
\cup{R_{k}^{*}} (B) &\subseteq 
\text{~~~[as $\cup{R_{k}} (B) \subseteq \cup{R_0}(\parent_{P_0}(B))$]}\\
\cup{R_0^*}(\parent_{P_0}(B))  &=\text{~~~[as $P_0 = \pr(\mu)$
  and $R_0^* = R_\mu^*=R_\mu$]}\\
\cup{R_\mu}(\parent_{\pr(\mu)}(B))  &= \text{~~~[by
  Lemma~\ref{prprop}~(ii), 
  $\tuple{\pr(\mu),R_\mu} = \tuple{\pr(\mud),R_{\mud}}$]}\\ 
\cup{R_\mud}(\parent_{\pr(\mud)}(B))  &=\text{~~~[by definition of $R_\mud$]}\\ 
\cup \{C\in \pr(\mud) ~|~ C \subseteq \mud(\parent_{\pr(\mud)}(B))\}
&= \text{~~~[by Lemma~\ref{prop}~(ii)]} \\
\mud(\parent_{\pr(\mud)}(B)) &= \text{~~~[by Lemma~\ref{prop}~(i)]} \\
\mud(B)&.
\end{align*}

\noindent
Thus, since, by Lemma~\ref{prprop}~(i), 
$P' \preceq \pr(\mu')$, by Lemma~\ref{prop}~(iv), 
$P' \preceq P_\mu = \pr(\mud)$ and both
$\mu'$ and $\mud$ are disjunctive, we have that 
for any $X\in \wp(\Sigma)$, 
\begin{align*} 
\mu'(X) &= \text{~~~~~[by Lemma~\ref{prop}~(iii)]}\\
\cup \{\mu'(B)~|~ B\in P',\: B\cap X\neq \varnothing\}   &\subseteq 
\text{~~~~~[as $\mu'(B)\subseteq \mud(B)$]}\\
\cup \{\mud(B)~|~ B\in P',\: B\cap X\neq \varnothing\} & = 
\text{~~~~~[by Lemma~\ref{prop}~(iii)]}\\ 
\mud(X) &\subseteq \text{~~~~~[as $\mud \sqsubseteq \mu$]}\\
\mu(X)&.
\end{align*}

\noindent
Thus, $\mu'$ is a refinement of $\mu$.  We have that
$P' \preceq \pr(\mu')$,
$R'=R_k$ is (as shown above) reflexive and because
$\tuple{P',R'}$ is the output 
partition-relation pair, for all $B,C\in P'$, if
$C\cap\pre(B)\not=\varnothing$ then
$\cup{R'}(C)\subseteq\pre(\cup{R'}(B))$. Hence, by Lemma~\ref{ltre} we
obtain that $\mu'$ is forward complete for $\pre$. Thus, $\mu'$ is a
disjunctive refinement of $\mu$ that is forward complete for $\pre$
so that $\mu'\sqsubseteq \cS_{\cup,\pre}(\mu)$.

\noindent
In order to conclude the proof, 
let us show that $\cS_{\cup,\pre}(\mu)\sqsubseteq \mu'$.
We first show by induction that for any $i\in [0,k]$ and $B\in P_i$,
we have that
$\cup{R_i} (B) \in \img(\cS_{\cup,\pre}(\mu))$:
\begin{description}
\item[($i=0$)] We have that $\tuple{P_0,R_0}=\tuple{P_\mu,R_\mu}$ so that for
any $B\in P_0$, by Lemma~\ref{prop}~(ii), 
$\cup{R_0}(B) = \cup \{C\in \pr(\mu)~|~ C
\subseteq \mu(B)\} = \mu(B)$. Hence, $\cup{R_0}(B) \in \img(\mu)
\subseteq \img(\cS_{\cup,\pre}(\mu))$. 
\item[($i+1$)] Let $C\in P_{i+1} = \splitt(P_i, \pre(\cup{R_i}(B_i)))$
for some $B_i \in P_i$. 
If $C\cap \pre(B_i) = \varnothing$ then,  
by lines 6-7, 
$\cup{R_{i+1}}(C) = \cup{R_i}(\parent_{P_i}(C))$ so that, by
inductive hypothesis, $\cup{R_{i+1}}(C) \in
\img(\cS_{\cup,\pre}(\mu))$. 
On the other hand, if $C\cap \pre(B_i) \neq \varnothing$ then, by
lines 6-7, $\cup{R_{i+1}}(C) = \cup{R_i}(\parent_{P_i}(C)) \cap 
\pre(\cup{R_i}(B_i))$. By inductive hypothesis, we have that
$\cup{R_i}(\parent_{P_i}(C)) \in
\img(\cS_{\cup,\pre}(\mu))$ and 
$\cup{R_i}(B_i) \in \img(\cS_{\cup,\pre}(\mu))$. 
Also, since $\cS_{\cup,\pre}(\mu)$ is 
forward complete for $\pre$, $\pre(\cup{R_i}(B_i)) 
\in \img(\cS_{\cup,\pre}(\mu))$. Hence, $\cup{R_{i+1}}(C) \in
\img(\cS_{\cup,\pre}(\mu))$. 
\end{description}

\noindent
As observed above, $R_k$ is reflexive so that for any $B\in P_k$, $B
\subseteq \cup{R_k}(B)$. For any $B\in P'$, we have that 
\begin{align*}
\cS_{\cup,\pre}(\mu)(B) &\subseteq  \text{~~~~~[as $B \subseteq
  \cup{R_k}(B)$]} \\
\cS_{\cup,\pre}(\mu)(\cup{R_k}(B)) &= \text{~~~~~[as
   $\cup{R_k}(B) \in \img(\cS_{\cup,\pre}(\mu))$]}\\
\cup{R_k}(B)  &\subseteq \text{~~~~~[as $R_k \subseteq {R_k}^{\!\!*}$]}\\
\cup{R_k^*} (B)  &= \text{~~~~~[by definition]}\\
\mu'(B)&. 
\end{align*}
Therefore, for any $X\in \wp(\Sigma)$, 
\begin{align*} 
\cS_{\cup,\pre}(\mu)(X) &\subseteq \text{~~~~~[as $X\subseteq
  \cup \{B\in P'~|~ B\cap X\neq \varnothing\}$]}\\
\cS_{\cup,\pre}(\mu)(\cup \{B\in P'~|~ B\cap X\neq
\varnothing\}) &= 
\text{~~~~~[as $\cS_{\cup,\pre}(\mu)$ is
  additive]} \\
\cup \{\cS_{\cup,\pre}(\mu)(B)~|~ B\in P',\: B\cap X\neq
\varnothing\}  &\subseteq 
\text{~~~~~[as $\cS_{\cup,\pre}(\mu)(B) \subseteq \mu'(B)$]}\\
\cup \{\mu'(B)~|~ B\in P',\: B\cap X\neq
\varnothing\} &= 
\text{~~~~~[as $\mu'$ is disjunctive, by Lemma~\ref{prop}~(iii)]} \\
\mu'(X)&. 
\end{align*}
We have therefore shown that $\cS_{\cup,\pre}(\mu)\sqsubseteq
\mu'$.
\end{proof}

Thus, $\BasicSA$ computes
the forward $\{\cup,\pre\}$-complete shell of any input abstract
domain. As a consequence, $\BasicSA$
allows us to compute both simulation relation and 
equivalence when $\mu_\ell$ is  the initial abstract domain.

\begin{corollary}\label{maincoro}
Let $\cK=(\Sigma,\sra,
\ell)$ be a finite Kripke structure and 
$\mu_\ell\in \uco(\wp(\Sigma))$ be the abstract domain induced
by $\ell$.  Then, $\BasicSA(\mu_\ell)= \tuple{P',R'}$ where
$P'=P_{\mathrm{sim}}$ and, for any $s_1,s_2\in \Sigma$, 
$(s_1,s_2)\in R_{\mathrm{sim}} \;\Lra\; (P_{\mathrm{sim}}(s_1), 
P_{\mathrm{sim}}(s_2))\in R'$.  
\end{corollary}
\begin{proof}
Let $\mu_\cK = \cS_{\cup,\pre}(\mu_\ell )$.
By Theorem~\ref{mainAlgo}, if
$\BasicSA(\mu_\ell)=\tuple{P',R'}$ then 
$\mu_{\tuple{P',R'}}=\mu_\cK$.
By Theorem~\ref{fcs}, 
$\pr(\mu_\cK) =P_{\mathrm{sim}}$. By Lemma~\ref{prprop}~(i), 
$P' \preceq \pr(\mu_{\tuple{P',R'}}) = \pr(\mu_\cK) =P_{\mathrm{sim}}$.   
It remains to show that 
$P_{\mathrm{sim}}= \pr(\mu_{\tuple{P',R'}}) \preceq P'$. 
Let $\{\tuple{P_i,R_i}\}_{i\in [0,k]}$ be the sequence of
partition-relation pairs computed by $\BasicSA$, where
$\tuple{P_0,R_0}= \tuple{P_{\mu_\ell},R_{\mu_\ell}}$ and 
$\tuple{P_k,R_k}=\tuple{P',R'}$.  
We 
show by induction that for any $i\in [0,k]$, 
we have that $\pr(\mu_{\tuple{P',R'}}) \preceq
P_i$. 
\begin{description}
\item[($i=0$)] Since $\mu_{\tuple{P',R'}} 
\sqsubseteq \mu_\ell$, we have that $\pr(\mu_{\tuple{P',R'}})
\preceq \pr(\mu_\ell) = P_0$. 
\item[($i+1$)] Consider $B\in \pr(\mu_{\tuple{P',R'}})$.
We have that $P_{i+1} = \splitt(P_i, \pre_\sra(\cup{R_i}(B_i)))$
for some $B_i \in P_i$. We have shown 
in the proof of Theorem~\ref{mainAlgo} that $\cup{R_i}(B_i) \in
\mu_\cK= \mu_{\tuple{P',R'}}$. Since $ \mu_{\tuple{P',R'}}$ 
is forward complete
for $\pre$, we also have that $\pre(\cup{R_i}(B_i))\in
\mu_{\tuple{P',R'}}$. Hence,  $B\cap
\pres(\cup{R_i}(B_i)) \in \{\varnothing, B\}$.
By inductive hypothesis, $\pr(\mu_{\tuple{P',R'}}) \preceq
P_i$ so that there
exists some $C\in P_i$ such that $B\subseteq C$. 
Since $P_{i+1} =
\splitt(P_i, \pre_\sra(\cup{R_i}(B_i)))$, note that if $C\cap
\pre_\sra(\cup{R_i}(B_i)) \neq \varnothing$ then 
$C\cap
\pre_\sra(\cup{R_i}(B_i)) \in P_{i+1}$ and if $C\smallsetminus 
(\pre_\sra(\cup{R_i}(B_i))) \neq \varnothing$ then 
$C\smallsetminus (\pre_\sra(\cup{R_i}(B_i)))\in P_{i+1}$. 
Moreover, if 
$B\cap
\pres(\cup{R_i}(B_i)) = \varnothing$ then $B \subseteq 
C\smallsetminus
(\pre_\sra(\cup{R_i}(B_i)))$, while if 
$B\cap
\pres(\cup{R_i}(B_i)) = B$ then
$B \subseteq C\cap
\pre_\sra(\cup{R_i}(B_i))$. In both cases, there exists some $D\in
P_{i+1}$ such that $B\subseteq D$.
\end{description}
Thus, $P'=\Psim$.
\\
The proof of Theorem~\ref{mainAlgo} 
shows that $R'$ is reflexive. Moreover, that proof also shows that for
any $B\in P'$, $\cup R'(B) \in \mu_\cK$. Then, for any $B\in P'$:
\begin{align*}
\cup  R'^{*} (B) &=  \text{~~~~~[by definition of $\mu_{\tuple{P',R'}}$]} \\
\mu_{\tuple{P',R'}}(B) &\subseteq \text{~~~~~[because $R'$ is
reflexive]} \\
\mu_{\tuple{P',R'}}(\cup R'(B))&=  \text{~~~~~[because
$\mu_{\tuple{P',R'}} =\mu_\cK$]} \\
\mu_\cK(\cup R'(B)) &= \text{~~~~~[because
$\cup R'(B) \in \mu_\cK$]} \\
\cup R'(B) & 
\end{align*}
and therefore $R'$ is transitive. Hence, 
for any $s_1,s_2\in \Sigma$,
\begin{align*}
(s_1,s_2)\in
R_{\mathrm{sim}} &\Lra  \text{~~~~~[by Theorem~\ref{fcs}]} \\
s_2 \in
\mu_\cK(\{s_1\}) & \Lra \text{~~~~~[because
  $\mu_\cK=\mu_{\tuple{P',R'}}$]} \\
s_2 \in
\mu_{\tuple{P',R'}}(\{s_1\})& \Lra \text{~~~~~[by definition of
  $\mu_{\tuple{P',R'}}$]} \\
(P'(s_1),P'(s_2)) \in R'^{*} & \Lra  \text{~~~~~[because
  $P' = \Psim$ and $R'^{*}=R'$]}\\
(\Psim(s_1),\Psim(s_2)) \in R'.& \qedhere
\end{align*}
\end{proof}

\linesnumbered
\begin{algorithm}[t]
\small
\SetVline
\Setnlskip{1.5em}
\SetAlTitleFnt{textsc}
\Indm
$\RefinedSA(\mathit{PartitionRelation} ~ \tuple{P,\mathit{Rel}})$ $\{$

\Indp

\lForAll{$B\in P$}{$\ppRel(B):=\Sigma$;}

\While{$\exists B \in P$ \KwSty{such that} $\pre(\cup\mathit{Rel}(B)) \neq \ppRel(B)$}{
\tcp{\textrm{~$\Inv_1$:} $\forall B \in P.\; \pre(\cup \mathit{Rel}(B))\subseteq \ppRel(B)$}

\tcp{\textrm{~$\Inv_2$:} $\forall B,C \in P.\; C\cap\pre(B)\neq\varnothing \,\Rightarrow \cup Rel(C)\subseteq\ppRel(B)$}
$\mathit{Remove}:=\ppRel(B)\smallsetminus \pre(\cup \mathit{Rel}(B))$\;
$\ppRel(B):=\pre(\cup \mathit{Rel}(B))$\;
$P_{\mathrm{prev}} := P$; ~$B_{\mathrm{prev}} :=B$\;
$P := \mathit{Split}(P,\ppRel(B))$\;
\ForAll{$C\in P$}{
	$\mathit{Rel}(C):=\{D\in P~|~D\subseteq\cup \mathit{Rel}(\parent_{P_{\mathrm{prev}}}(C))\}$\;
	\lIf{$C\in P\smallsetminus
          P_{\mathrm{prev}}$}{$\ppRel(C):= \ppRel(\parent_{P_{\mathrm{prev}}}(C))$}\;
}
\ForAll{$C\in P$  \KwSty{such that} $C\cap\pre(B_{\mathrm{prev}})\neq\varnothing$}{
$Rel(C):=\{D\in Rel(C) ~|~ D\cap \mathit{Remove}=\varnothing\}$\;
}
}
\Indm
$\}$
\caption{Refined Simulation Algorithm.}\label{simequiv2}
\end{algorithm}

\subsection{Refining the Algorithm}
The $\BasicSA$ algorithm is refined to
the $\RefinedSA$ procedure in Figure~\ref{simequiv2}.  
This is obtained by adapting the ideas of Henzinger et al.'s
$\mathit{RefinedSimilarity}$ procedure 
in Figure~\ref{hhkfig} to our $\BasicSA$ algorithm. The
following points
show that 
this algorithm $\RefinedSA$ remains correct, i.e.\
the input-output behaviours of $\BasicSA$ and 
$\RefinedSA$ are the same. 

\begin{itemize}

\item[--] For any block $B$ of the current partition $P$, 
 the predecessors of the blocks in the ``previous'' 
relation $\mathit{Rel}_{\mathrm{prev}}(B)$ 
are maintained as a set $\ppRel(B)$. 
Initially, at line~2, $\ppRel(B)$ is set to contain all the states in $\Sigma$. 
Then, when a block $B$ is
selected by the while-loop 
at some iteration $i$,  $\ppRel(B)$ is updated at line~7 in order to save
the states in $\pre(\cup\!\Rel(B))$ at this iteration $i$.

\item[--] If $C$ is a newly generated block  after splitting $P$ and
  $\parent_{P_{\mathrm{prev}}}(C)$  is
  its corresponding parent block in the partition before splitting then
  $\ppRel(C)$ is set at line~12
  as  $\ppRel(\ok{\parent_{P_{\mathrm{prev}}}}(C))$.
  Therefore, since the current relation $\Rel$ decreases
  only~---~i.e., if $i$ and $j$ are iterations such that $j$ follows
  $i$ and $B,B'$ are blocks such that $B'\subseteq B$ then
  $\cup\!\Rel_j(B') \subseteq \cup\!\Rel_i(B)$~---~at 
each iteration, the following invariant
  $\Inv_1$ holds:  for any block
  $B\in P$, $\pre(\cup\!\Rel(B))
  \,\subseteq\, \ppRel(B)$. Initially, 
$\Inv_1$ is satisfied because for any block
$B$, $\ppRel(B)$ is initialized to $\Sigma$ at line~2.

\item[--]
The crucial point is the invariant $\Inv_2$: if
$C\sra^{\exists\exists} B$ and $D\in \Rel(C)$ then $D\subseteq
\ppRel(B)$. Initially, 
this invariant property is clearly satisfied because for any block
$B$, $\ppRel(B)$ is initialized to $\Sigma$. 
Morever, $\Inv_2$ is maintained at each iteration because at line~6
$\Remove$ is set to 
$\ppRel(B) \smallsetminus \pre(\cup\!
\Rel(B))$ and for any block $C$ such that $C\sra^{\exists\exists}
B_{\mathrm{prev}}$ if some block $D$ is contained in $\Remove$ then
$D$ is removed from $\Rel(C)$ at line~14. 

\end{itemize}

Thus, if the exit condition of the while-loop of $\RefinedSA$ is
satisfied then, by invariant $\Inv_2$,  the exit condition of
$\BasicSA$ is satisfied as well. 

Finally, let us remark that the exit condition of the
 while-loop, namely $\forall B\in P.\: 
\pre(\cup\mathit{Rel}(B)) = \ppRel(B)$, is strictly
weaker than the exit condition that we would obtain as counterpart of
the exit condition of the while-loop of  Henzinger et al.'s 
$\mathit{RefinedSimilarity}$ procedure,
i.e.\ $\forall B\in P.\: 
\mathit{Rel}(B) = \mathit{Rel}_{\mathrm{prev}}(B)$.


\linesnumbered
\begin{algorithm}[Ht]
\small
\SetVline
\Setnlskip{1.5em}
\SetAlTitleFnt{textsc}
\Indm
$\SA(\mathit{PartitionRelation} ~ \tuple{P,\mathit{Rel}})$ $\{$

\Indp

\tcp{\lForAll{$B\in P$}{$\ppRel(B):=\Sigma;$}}
\lForAll{$B\in P$}{$\mathit{Remove}(B):=\Sigma\smallsetminus\pre(\cup\mathit{Rel}(B))$;}

\While{$\exists B \in P$ \KwSty{such that} $\mathit{Remove}(B) \neq \varnothing$}{
\tcp{\textrm{~$\Inv_3$:} $\forall C \in P.\; \mathit{Remove}(C)=
  \ppRel (C) \smallsetminus \pre(\cup\mathit{Rel}(C))$}

\tcp{\textrm{~$\Inv_4$:} $\forall C \in P.\; \mathit{Split}(P,\ppRel(C))=P$}

\tcp{$\ppRel(B) := \pre(\cup \mathit{Rel}(B));$}

$\mathit{Remove}:=\mathit{Remove}(B)$\;
$\mathit{Remove}(B):= \varnothing$\;
$B_{\mathrm{prev}} :=B$\;
$P_{\mathrm{prev}} := P$\;
$P := \mathit{Split}(P,\mathit{Remove})$\;
\ForAll{$C\in P$}{
	$\mathit{Rel}(C):=\{D\in P~|~D\subseteq\cup \mathit{Rel}(\ok{\parent_{P_{\mathrm{prev}}}(C)})\}$\;
	\If{$C\in P\smallsetminus P_{\mathrm{prev}}$}{
        $\mathit{Remove}(C):= \mathit{Remove}(\ok{\parent_{P_{\mathrm{prev}}}}(C))$\;
	\tcp{$\ppRel(C):= \ppRel(\ok{\parent_{P_{\mathrm{prev}}}}(C));$}
        }
}
$\mathit{RemoveList} := \{D\in P~|~D\subseteq \mathit{Remove}\}$\;
\ForAll{$C\in P$  \KwSty{such that} $C\cap\pre(\ok{B_{\mathrm{prev}}})\neq\varnothing$}{
	\ForAll{$D\in \mathit{RemoveList}$}{
		\If{$D\in \mathit{Rel}(C)$}{
			$\mathit{Rel}(C):=\mathit{Rel}(C)\smallsetminus\{D\}$\;
			\ForAll{$s \in \pre(D)$ \KwSty{such that} $s\not\in \pre(\cup\mathit{Rel}(C))$}{
			$\mathit{Remove}(C):=\mathit{Remove}(C)\cup\{s\}$\;
				
			}
		}
	}
}
}
\Indm
$\}$
\caption{The Simulation Algorithm $\SA$.}\label{simequiv3}
\end{algorithm}

\subsection{The Final Algorithm}
Following the underlying ideas that lead from 
$\mathit{RefinedSimilarity}$ to $\HHK$, the
algorithm $\RefinedSA$ is further refined to its final version
$\SA$ in Figure~\ref{simequiv3}.  The idea is that instead of
recomputing at each iteration of the while-loop the set
$\mathit{Remove} = \ppRel(B)\smallsetminus
\pre(\cup\mathit{Rel}(B))$ for the selected block $B$, we maintain a set
of states $\Remove(B)\subseteq \Sigma$ for each block $B$ of the
current partition. For any block $C$, $\Remove(C)$ is updated in order
to satisfy the invariant condition $\Inv_3$: $\Remove(C)$ contains
exactly the set of states that belong to $\ppRel(C)$ but
are not in $\pre(\cup\!\Rel(C))$, where 
$\ppRel(C)$ is logically defined as in
$\RefinedSA$ but is not really stored. Moreover, the invariant
condition $\Inv_4$ ensures that, for any block $C$,
$\ppRel(C)$ is a union of blocks of the current
partition.  This allows us to replace the
operation $\Split(P,\pre(\cup \! \Rel(B)))$ in $\RefinedSA$
with the equivalent split operation
$\Split(P,\Remove)$. The correctness of such replacement  
follows from the invariant condition 
$\Inv_4$  by exploiting the following
general remark. 
\begin{lemma}\label{lemparti}
  Let $P$ be a partition, $T$ be a union of blocks in $P$ and $S\subseteq T$.
  Then, $\Split(P,S)=\Split(P,T\smallsetminus S)$.  
\end{lemma}
\begin{proof}
Assume that $B\cap T=\varnothing$, so that $B\cap S=\varnothing$.  Then, 
    $$B\cap (T\smallsetminus S) = B\cap (T\cap \neg S) = \varnothing =
    B\cap S$$
and 
$$B\smallsetminus (T\smallsetminus S) = (B \cap \neg T) \cup (B\cap S)
= B = B\smallsetminus S$$
so that $B$ is split neither by $T\smallsetminus S$ nor by $S$. 
\\
Otherwise, if $B\cap T\neq \varnothing$, because $T$ is a union of
blocks, then  $B\subseteq T$. Then,
$$B\cap (T\smallsetminus S) = B\cap (T\cap \neg S) =
B\cap \neg S = B\smallsetminus S$$
and 
$$B\smallsetminus (T\smallsetminus S) = (B \cap \neg T) \cup (B\cap S)
= B \cap S$$
so that 
$B$ is split by $T\smallsetminus S$ into $B_1$ and $B_2$ 
if and only if 
$B$ is split by $S$ into $B_1$ and $B_2$. 
We have thus shown that $\Split(P,S)=\Split(P,T\smallsetminus S)$.  
\end{proof}
 
The equivalence between $\SA$ and $\RefinedSA$ 
is a consequence of the following
observations.
\begin{itemize}

\item[--] 
  Initially, the invariant properties
  $\Inv_3$  and $\Inv_4$ clearly hold because for any block $B$,
  $\ppRel(B) =\Sigma$.  

\item[--] When a block $B_{\mathrm{prev}}$ of the current partition is
  selected by the while-loop, the corresponding remove set
  $\Remove(B_{\mathrm{prev}})$ is set to empty at line~9.  The
  invariant $\Inv_3$, namely $\forall C.\;
  \mathit{Remove}(C)=\ppRel(C)\smallsetminus
  \pre(\cup\mathit{Rel}(C))$, is maintained at each iteration because
  for any block $C$ such that $C\sra^{\exists\exists}
  B_{\mathrm{prev}}$ the for-loop at lines~23-24 incrementally adds to
  $\Remove(C)$ all the states $s$ that are in
  $\ppRel(C)$ but not in $\pre(\cup\!\Rel(C))$.

\item[--] If $C$ is a newly generated block after splitting $P$ and
  $\parent_{P_{\mathrm{prev}}}(C)$   is
  its corresponding parent block in the partition before splitting then
  $\Remove(C)$ is set to $\Remove(\parent_{P_{\mathrm{prev}}}(C))$ by
  the for-loop at lines 13-17.

\item[--] As in $\RefinedSA$, 
for any block $C$ such that $C\sra^{\exists\exists}
B_{\mathrm{prev}}$, all the  blocks  that are contained in 
$\Remove(B_{\mathrm{prev}})$
are removed from $\Rel(C)$ by the for-loop at lines~20-22.
\end{itemize}

If the exit condition of the while-loop of $\SA$ is satisfied
then, by $\Inv_1$ and $\Inv_3$, the exit condition of $\RefinedSA$ is
satisfied as well.

\section{Complexity}

\subsection{Data Structures}\label{ds}
$\SA$ is implemented by using the following data structures.

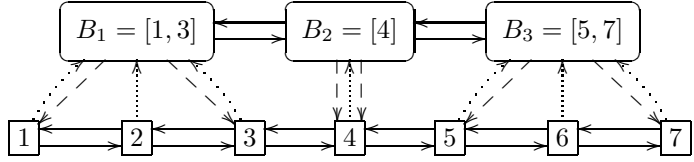
\begin{figure}
\begin{center}
\mbox{\xymatrix@C=8pt@R=8.5pt{
 &*++[F-:<3pt>]{B_1=[1,3]}\ar@{-->}[ldd]\ar@{-->}[rdd]\ar@<-0.7ex>[rr] & &
 *++[F-:<3pt>]{B_2=[4]}\ar@<1ex>@{-->}[dd]\ar@<-1ex>@{-->}[dd]\ar@<-0.7ex>[rr]
\ar@<-0.7ex>[ll]
 & 
 &*++[F-:<3pt>]{B_3=[5,7]}\ar@{-->}[rdd]\ar@{-->}[ldd]\ar@<-0.7ex>[ll] &  \\
&&&&&&\\
*+[F]{1}\ar@<-0.7ex>[r]\ar@/^/@{.>}[uur] &
*+[F]{2}\ar@<-0.7ex>[r]\ar@<-0.7ex>[l] \ar@{.>}[uu]  & 
*+[F]{3}\ar@<-0.7ex>[r]\ar@<-0.7ex>[l]  \ar@/_/@{.>}[uul] & 
*+[F]{4}\ar@<-0.7ex>[r]\ar@<-0.7ex>[l] \ar@{.>}[uu]  & 
*+[F]{5}\ar@<-0.7ex>[r]\ar@<-0.7ex>[l] \ar@/^/@{.>}[uur] & 
*+[F]{6}\ar@<-0.7ex>[r]\ar@<-0.7ex>[l] \ar@{.>}[uu] &
*+[F]{7}\ar@<-0.7ex>[l] \ar@/_/@{.>}[uul] \\ 
    }
  }
\end{center}
\caption{Partition representation.}\label{partitionFig}
\end{figure}

\begin{itemize}

\item[{\rm (i)}] The set of states $\Sigma$ is represented as a doubly
  linked list where 
 each state $s\in \Sigma$ (represented as an integer) stores
  the list of its
  predecessors in $\pre(\{s\})$. This provides a representation of
  the input transition system. Any state $s\in \Sigma$ also stores a
  pointer to the block of the current partition that contains $s$.

\item[{\rm (ii)}] The states of any block $B$ of the current partition
  are consecutive in the list $\Sigma$, so that $B$ is represented by
  a record that contains two pointers to the first and to the last state in
  $B$ (see Figure~\ref{partitionFig}).  This structure allows us to
  move a state from a block to a different block in constant time. 
  Moreover, any block $B$ stores its corresponding remove set
  $B.\!\Remove$, which is represented as a list of (pointers to)
  states.

\item[{\rm (iii)}] Any block $B$ additionally stores an integer
  array $\mathit{RelCount}$ 
  that is indexed over $\Sigma$ and is defined as follows: for any $x\in \Sigma$,
  $B.\mathit{RelCount}(x) = \sum_{C \in \Rel(B)}
  |\{(x,y)~|~x\sra y,\: y\!\in\! C\}|$
  is the number of transitions from $x$ to some block $C\in \Rel(B)$.
   The array $\mathit{RelCount}$ allows to
  implement in constant time the test $s\not\in \pre(\cup\!\Rel(C))$
  at line~23 as $C.\mathit{RelCount}(s)=0$.

\item[{\rm (iv)}] The current partition is stored as a doubly linked
  list $P$ of blocks. Newly generated blocks are appended or prepended
  to this
  list.  Blocks are scanned from the beginning of this list by
  checking whether the corresponding remove set is empty or
  not. If an empty remove set of some block $B$ becomes nonempty then
  $B$ is moved to the end of $P$.
   
\item[{\rm (v)}] The current relation $\Rel$ on the current partition
  $P$ is stored as a resizable $|P|\times |P|$ boolean matrix
  \cite[Section~17.4]{cormen}. The
  algorithm adds a new entry to this matrix, namely a new row and a
  new column, as long as a block $B$ is
  split at line~12 into two new blocks $B\smallsetminus \Remove$ and
  $B\cap \Remove$: the new block $B\smallsetminus \Remove$ replaces
  the old block $B$ in $P$ while a new entry in the matrix $\Rel$
  corresponds to the new block $B\cap \Remove$.  We will observe later
  that the overall number of newly generated blocks by the splitting
  operation at line~12 is exactly given by
  $2(|P_{\mathrm{sim}}|-|P_{\mathrm{in}}|)$.  Hence, the total number
  of insert operations in the matrix $\Rel$ is
  $|P_{\mathrm{sim}}|-|P_{\mathrm{in}}|\leq |\Psim|$. Since an insert
  operation in a resizable array (whose capacity is doubled as needed)
  takes an amortized constant time, the overall cost of inserting new
  entries to the matrix $\Rel$ is in
  $O(|P_{\mathrm{sim}}|^2)$-time. Let us recall that the standard C++
  vector class implements a resizable array so that a resizable
  boolean matrix can be easily implemented as a C++ vector of boolean
  vectors: in this implementation, the algorithm adds a new entry
  to a $N\times N$ matrix by first inserting a new vector of size
  $N+1$ containing 
  $\mathit{false}$ values and then by inserting $N+1$ $\mathit{false}$
  values in the $N+1$ boolean vectors.    
\end{itemize}

\subsection{Space and Time Complexity}
Let $B\in P_{\mathrm{in}}$ be some block of
the initial partition $ P_{\mathrm{in}}$ 
and let $\langle B_i\rangle_{i\in \mathit{It}}$ be the
sequence of all the blocks selected by the while-loop in a sequence
$\mathit{It}$ of
iterations such that:
\begin{itemize} 
\item[(a)] for any $i\in \mathit{It}$,
$B_i\subseteq B$; 
\item[(b)] if an iteration $j\in \mathit{It}$ follows an iteration
$i\in \mathit{It}$, denoted by $i<j$, then $B_j$ is contained in
$B_i$.
\end{itemize}
 
\noindent
Observe that $B$ is the parent block in $P_{\mathrm{in}}$ of all the
$B_i$'s.  Then, one key property of the $\SA$ algorithm is
that the remove sets in $\{\Remove(B_i)\}_{i\in \mathit{It}}$ are
pairwise disjoint so that $\sum_{i\in \mathit{It}} |\Remove(B_i)| \leq
|\Sigma|$.  This property guarantees that if the test $D\in
\mathit{RemoveList}$ at line~20 is positive at some iteration $i\in
\mathit{It}$ then for any block $D'\subseteq D$ and for any successive
iteration $j>i$, with $j\in \mathit{It}$, the test $D' \in
\mathit{RemoveList}$ will be negative.  Moreover, if the test $D\in \Rel(C)$
at line~21 is positive at some iteration $i\in \mathit{It}$, so that
$D$ is removed from $\Rel(C)$, then for all the blocks $D'$ and $C'$
such that $D'\subseteq D$ and $C'\subseteq C$ the test $D'\in
\Rel(C')$ will be negative for all the iterations $j>i$.  As a further
consequence, since a splitting operation $\Split(P,\Remove)$ can be
executed in $O(|\mathit{Remove}|)$-time, it turns out that the overall cost of
all the splitting operations is in
$O(|P_{\mathrm{sim}}||\Sigma|)$-time.  Furthermore, by using the data
structures described by points~(iii) and~(v) in 
Section~\ref{ds}, the tests $D\in \Rel(C)$
at line~21 and $s\not\in \pre(\cup\!\Rel(C))$ at line~23 can be
executed in constant time.
A careful analysis that exploits these key facts allows us to
show that 
the total running time of $\SA$ is in $O(|P_{\mathrm{sim}}||\sra|)$.

\begin{theorem}\label{complexity}
  The algorithm $\SA$ runs in 
  $O(|P_{\mathrm{sim}}||\sra|)$-time and 
$O(|P_{\mathrm{sim}}||\Sigma|\log|\Sigma|)$-space.
\end{theorem}
\begin{proof}
Let $\It$ denote the sequence of iterations of the while-loop for some run of
$\SA$, where for any $i,j\in \It$, $i<j$ means that $j$
follows 
$i$. Moreover, for any $i\in \It$, $B_i$ denotes the block selected by
the while-loop at
line~4, $\Remove(B_i)\neq \varnothing $ denotes the corresponding
nonempty remove set, $\pre(\cup\!\Rel (B_i))$ denotes the corresponding set
for $B_i$, while
$\tuple{P_i,\Rel_i}$ denotes 
the partition-relation pair at the entry point of the for-loop at line~19. 

\noindent
Consider the set $\cB\ud \{B_i\in P_i ~|~i\in \It\}$ of selected
blocks and the following relation on $\cB$: 
$$B_i \trianglelefteq B_j \;\:\Lra\;\: B_i \subsetneq B_j \text{~or~}
(B_i=B_j \;\&\;\, i\geq j)$$
It turns out that $\tuple{\cB,\tle}$ is a poset. In fact, $\tle$ is
trivially reflexive. Also, $\tle$ is transitive: assume that $B_i \tle B_j$ and $B_j \tle B_k$;
if $B_i=B_j=B_k$ then $i\geq j \geq k$ so that $B_i \tle B_k$;
otherwise either $B_i \subsetneq B_j$ or $B_j \subsetneq B_k$ so that
$B_i \subsetneq B_k$ and therefore $B_i \tle B_k$. Finally, $\tle$ is
antisymmetric: if $B_i \tle B_j$ and $B_j \tle B_i$ then $B_i=B_j$ and
$i\geq j \geq i$ so that $i=j$. Moreover, $B_i \triangleleft B_j$
denotes the corresponding strict order: this
happens when either $B_i \subsetneq B_j$ or $B_i = B_j$ and $i>j$.  

\noindent
The time complexity bound is shown  incrementally by the
following points.

\begin{itemize}

\item[{\rm (A)}]
For any $B_i,B_j\in \cB$, if $B_i\subseteq B_j$ and $j<i$ then  $\Remove(B_i)\cap
  \Remove(B_j)=\varnothing$. 

\noindent
\textit{Proof.} 
By invariant $\Inv_3$, $\Remove (B_j) \cap \pre(\cup\!
\Rel_j (B_j))=\varnothing$.  At
iteration $j$, $\Remove(B_j)$ is set to $\varnothing$ at
line~9. If $B_j$ generates, by the splitting operation at line~12, two
new blocks $B_1,B_2\subseteq B_j$ then their remove sets are set to
$\varnothing$ at line~16. 
Successively, $\SA$ may add at line~24 of some iteration
$k\geq j$  
a state $s$ to the remove set $\Remove(C)$ of 
a block $C\subseteq B_j$  only if $s\in \pre(\cup\!\Rel_k(C))$. 
We also have that $\cup\!\Rel_k(C)\subseteq \cup\!\Rel_j(B_j)$ so
that $\pre(\cup\!\Rel_k(C)) \subseteq \pre(\cup\!\Rel_j(B_j))$. 
Thus, if $B_i\subseteq B_j$ and $i>j$ then $\Remove(B_i)\subseteq
\pre(\cup \!\Rel_j(B_j))$. Therefore, 
$\Remove(B_j)\cap
  \Remove(B_i) \subseteq \Remove(B_j) \cap 
\pre(\cup\!\Rel_j(B_j)) = \varnothing$.

\item[{\rm (B)}] The
overall number of newly generated blocks by the splitting operation at
line~12 is 
$2(|P_{\mathrm{sim}}|-|P_{\mathrm{in}}|)$.

\noindent
\textit{Proof.} Let $\{P_i\}_{i\in [0,n]}$ be the sequence of
partitions computed by $\SA$ where $P_0$ is the initial
partition $P_{\mathrm{in}}$, 
$P_n$ is the final partition $\Psim$ and for all $i\in
[0,n-1]$, $P_{i+1} \preceq P_{i}$. The number of newly generated blocks
by one splitting operation that refines $P_i$ to $P_{i+1}$ is given by
$2(|P_{i+1}| - |P_i|)$. Thus, the
overall number of newly generated blocks is $\sum_{i=0}^{n-1}
2(|P_{i+1}| - |P_i|) =  
2(|P_{\mathrm{sim}}|-|P_{\mathrm{in}}|)$.

\item[{\rm (C)}] The time complexity of the for-loop at line~3 is in
$O(|P_{\mathrm{in}}||\sra|)$.

\noindent
\textit{Proof.} For any $B\in P_{\mathrm{in}}$, $\pre(\cup\!\Rel(B))$
is computed in $O(|\sra|)$-time,  so that $\Sigma\smallsetminus 
\pre(\cup\!\Rel(B))$
is computed in $O(|\sra|)$-time as well.
The time complexity of  the initialization of the remove sets
is therefore in $O(|P_{\mathrm{in}}||\sra|)$. 

\item[{\rm (D)}] The overall time complexity of lines~8 and~18 is in
$O(|P_{\mathrm{sim}}||\Sigma|)$. 

\noindent
\textit{Proof.}
Note that at line~18, $\Remove$ is a union of blocks of the current
partition $P$. As described in Section~\ref{ds}~(i), 
each state $s$ also stores a pointer to the 
block of the current partition that contains $s$.
The list of blocks $\mathit{RemoveList}$ is therefore 
computed by scanning all
the states in $\Remove(B_i)$, where $B_i$ is the selected block at
iteration $i$, so that
the overall time complexity of lines~8 and~18 
is bounded by $\textstyle 2\sum_{i\in \It}
|\Remove(B_i)|$.
For any block $E\in P_{\mathrm{sim}}$ of the final partition we
define the following subset of iterations:
$$\It_E \ud \{ i\in \It ~|~E \subseteq B_i\}.$$ 
Since for any
$i\in \It$, $P_{\mathrm{sim}}\preceq P_i$, we have that for any
$i\in \It$ there exists some $E\in P_{\mathrm{sim}}$ such that $i\in
\It_E$. Note that if $i,j\in
\It_E$ and $i < j$ then $B_j\subseteq B_i$ and, by point~(A), this
implies that $\Remove(B_i) \cap \Remove(B_j)=\varnothing$. 
Thus,
\begin{align*}
\textstyle
2 \sum_{i\in \It} |\Remove(B_i)| & \leq 
\text{~~~~~[by definition of $\It_E$]}\\
\textstyle
2 \sum_{E\in \Psim} \sum_{i\in \It_E} |\Remove(B_i)|  &\leq
\text{~~~~~[as the sets in $\{\Remove(B_i)\}_{i\in {\It_E}}$ are pairwise
  disjoint]}\\
\textstyle
2\sum_{E\in \Psim} |\Sigma| & = \\
2|\Psim||\Sigma|&. 
\end{align*}

\begin{figure*}[t]
\begin{center}
{
\scriptsize
\begin{codenumber}[frame=single] 
ListOfBlocks $\mathit[Split]$(PartitionRelation& P, SetOfStates S) { 
   ListOfBlocks split = empty;
   forall s in S do { 
      Block B = s.block;
      if (B.intersection == NULL) then {
         B.intersection = new Block;
         if (B.remove == $\;\varnothing$) then P.prepend(B.intersection);
                             else P.append(B.intersection);
         split.append(B);
      }
      move s from B to B.intersection;
      if (B == empty) then { 
         B = copy(B.intersection); 
         P.remove(B.intersection);
         delete B.intersection;
         split.remove(B);
      }
   }
   return split;
}

$\mathit[SplittingProcedure]$(P,S) {
   /* P$_\mathrm[prev]$ = P; */
   ListOfBlocks split = $\mathit[Split]$(P,S);
   /* $\mathit[assert]$(split == {B$\smallsetminus$S $\in$ P | B$\smallsetminus$S $\not\in$ P$_\mathrm[prev]$}) */
   forall B in split do { 
      Rel.addNewEntry(B.intersection);
      B.intersection.Remove = copy(B.Remove);
   }
   forall B in P do 
      forall C in split do Rel(B,C.intersection) = Rel(B,C);
   forall B in split do {
      forall C in P do Rel(B.intersection,C) = Rel(B,C);
      forall x in $\Sigma$ do B.intersection.RelCount(x) = B.RelCount(x);
   }
}
\end{codenumber}
}
\end{center}
\caption{C++ Pseudocode Implementation of the Splitting Procedure.}\label{fi}
\end{figure*}

\item[{\rm (E)}] The overall time complexity of line~10, i.e.\ of 
copying the list of states of the selected block $B$,
  is in
$O(|P_{\mathrm{sim}}||\Sigma|)$. 

\noindent
\textit{Proof.} 
For any block $E\in P_{\mathrm{sim}}$ of the final partition we
define the following subset of iterations:
$$\It_E \ud \{ i\in \It ~|~E \subseteq \Remove(B_i)\}.$$ 
Since for any
$i\in \It$, $P_{\mathrm{sim}}\preceq P_i$ and $\Remove(B_i)$ is a
union of blocks of $P_i$, it turns out that for any
$i\in \It$ there exists some $E\in P_{\mathrm{sim}}$ such that $i\in
\It_E$. Note that if $i,j\in
\It_E$ and $i \neq j$ then $B_j\cap B_i=\varnothing$: this is a
consequence of point~(A) because
$E\subseteq
\Remove(B_i)\cap \Remove(B_j)\neq \varnothing$ implies that
$B_j\not\subseteq B_i$ and $B_i\not\subseteq B_j$ so that $B_i \cap
B_j=\varnothing$. 
Thus,
\begin{align*}
\textstyle
\sum_{i\in \It} |B_i| & \leq 
\text{~~~~~[by definition of $\It_E$]}\\
\textstyle
\sum_{E\in \Psim} \sum_{i\in \It_E} |B_i|  &\leq
\text{~~~~~[as the blocks in $\{B_i\}_{i\in {\It_E}}$ are pairwise
  disjoint]}\\
\textstyle
\sum_{E\in \Psim} |\Sigma| & = \\
|\Psim||\Sigma|&. 
\end{align*}

\item[{\rm (F)}] The overall time complexity of lines~11-17 is in
$O(|P_{\mathrm{sim}}||\sra|)$. 

\noindent
\textit{Proof.}
Figure~\ref{fi} describes a C++ pseudocode implementation of
lines 11-17. By using the data structures described in
Section~\ref{ds}, and in particular in Figure~\ref{partitionFig}, all
the operations of the procedure $\mathit{Split}$ take constant time so
that any call $\mathit{Split}(P,S)$ takes $O(|S|)$ time. 
Let us now consider $\mathit{SplittingProcedure}$. 
\begin{itemize}
\item[--] The overall time complexity of the splitting operation at
  line~24 is in $O(|\Psim||\Sigma|)$. Each call
  $\mathit{Split}(P,\Remove(B_i))$ takes $O(|\Remove(B_i)|)$ time. Then, 
analogously to the proof of point~(D), the overall time complexity of
line~24 is bounded by $\sum_{i\in \It} |\Remove(B_i)| \leq
|\Psim||\Sigma|$. 
\item[--] The overall time complexity of the for-loop at lines~26-29 is
  in $O(|\Psim||\Sigma|)$. It is only worth noticing that since the
  boolean matrix that stores $\Rel$ is resizable, each operation at
  line~27 that adds a new entry to this resizable matrix has an
  amortized cost in $O(|\Psim|)$: in fact, the resizable 
matrix is just a resizable array $A$ of resizable arrays so that when we
add a new entry we need to add a new entry to $A$ and then a new entry
to each array in $A$ (cf.\ point~(v) in Section~\ref{ds}). 
Thus, the overall time complexity of line~26 is
in $O(|\Psim|^2)$.  
\item[--] The overall time complexity of the for-loop at lines~30-31 is
  in $O(|\Psim|^2)$.
\item[--] The overall time complexity of the for-loop at lines~32-35 is
  in $O(|\Psim||\sra|)$. This is a consequence of the
  fact that the overall time complexity of the for-loops at lines~33
  and~34 is in $O(|\Psim||\sra|)$.
\end{itemize}
Thus, the overall time complexity of 
$\mathit{SplittingProcedure}(P,\Remove)$ 
is in
$O(|P_{\mathrm{sim}}||\sra|)$.

\item[{\rm (G)}] The overall time complexity of lines~19-21 is in
$O(|P_{\mathrm{sim}}||\sra|)$. 

\noindent
\textit{Proof.} 
For any $B_i\in \cB$, let $\arr(B_i) \ud \sum_{x\in B_i}
|\pre(\{x\})|$ denote the number of transitions that end in some state
of $B_i$ and $\rem(B_i) \ud |\{ D\in P_i~|~ D\subseteq
\Remove(B_i)\}|$ denote the number of blocks of $P_i$ contained in
$\Remove(B_i)$. We also define two functions
$f_\vtl,f_\tle :\cB \ra \wp(\Psim)$  as follows:
\begin{align*}
f_\vtl (B_i) \ud \{D\in \Psim ~|~ D\cap (\cup
\{\Remove(B_j)~|~B_j\in \cB,\:  B_i
\vartriangleleft B_j\})=\varnothing\}\\
f_\tle (B_i) \ud \{D\in \Psim ~|~ D\cap (\cup
\{\Remove(B_j)~|~B_j\in \cB,\:  B_i
\tle B_j\})=\varnothing\}
\end{align*}
Let us show the following property:
$$\forall B_i\in \cB.\; \rem(B_i) + |f_\tle (B_i)| 
\leq |f_\vtl (B_i)|.\eqno(\ddagger)$$  
We first observe that since $\Psim \preceq P_i$,
$\rem(B_i) \leq |\{ D\in \Psim~|~D\subseteq
\Remove(B_i)\}|$. Moreover, the sets $\{ D\in \Psim~|~D\subseteq
\Remove(B_i)\}$ and $f_\tle (B_i)$ are disjoint and their union 
gives  $f_\vtl (B_i)$. Hence, 
\begin{align*}
\rem(B_i) + |f_\tle (B_i)| & \leq\\
|\{ D\in \Psim~|~D\subseteq
\Remove(B_i)\}| + |f_\tle (B_i)| & = \\
| \{ D\in \Psim~|~D\subseteq
\Remove(B_i)\} \cup f_\tle (B_i)| &= \\
 |f_\vtl (B_i)| &.
\end{align*}

Given, $B_k\in \cB$, let us show by induction on the height
$h(B_k)\geq 0$
of $B_k$ in the poset $\tuple{\cB,\tle}$ that
$$\textstyle \sum_{B_i \tle B_k} \arr(B_i) \rem(B_i) \leq
\arr(B_k)|f_\vtl(B_k)|.\eqno(*)$$

\medskip
\noindent
($h(B_k)=0$): By property~$(\ddagger)$, $\rem(B_k) \leq |f_\vtl(B_k)|$ so that 
$$\textstyle \sum_{B_i \tle B_k} \arr(B_i) \rem(B_i) =
\arr(B_k)\rem(B_k) \leq \arr(B_k) |f_\vtl(B_k)|.$$

\medskip
\noindent
($h(B_k)>0$): Let $\max(\{B_i\in \cB ~|~ B_i \vartriangleleft
  B_k\}) = \{C_1,...,C_n\}$. Note that if $i\neq j$ then $C_i\cap
  C_j=\varnothing$, so that $\sum_i \arr(C_i) \leq \arr (B_k)$, 
since $\cup_i C_i \subseteq B_k$. 
Let us observe that for any maximal $C_i$, $f_\vtl(C_i) \subseteq
f_\tle (B_k)$ because $\cup \{\Remove(B_j)~|~B_j\in \cB,\: 
B_k \tle B_j \} \subseteq \cup \{\Remove(B_j)~|~B_j \in \cB,\: 
C_i \vtl B_j \}$ since $B_k \tle B_j$ and $C_i \vtl B_k$ 
imply $C_i \vtl B_j$. 

Hence, we have that
\begin{align*}
\textstyle
\sum_{B_i\tle B_k} \arr(B_i) \rem(B_i)  & = 
\text{~~[by maximality of $C_i$'s]}\\
\textstyle
\arr(B_k)\rem(B_k) + \sum_{C_i} \sum_{D\tle C_i} \arr(D) \rem(D)  & \leq
\text{~~[by inductive hypothesis on $h(C_i)<h(B_k)$]}\\
\textstyle
\arr(B_k)\rem(B_k) + \sum_{C_i} \arr(C_i) |f_\vtl(C_i)| & \leq 
\text{~~[as $f_\vtl(C_i) \subseteq f_\tle (B_k)$]}\\ 
\textstyle
\arr(B_k)\rem(B_k) + |f_\tle(B_k)| \sum_{C_i} \arr(C_i) & \leq 
\text{~~[as $\textstyle\sum_{C_i} \arr(C_i) \leq \arr (B_k)$]}\\
\textstyle
\arr(B_k)\rem(B_k) + |f_\tle(B_k)| \arr(B_k)  & = \\
\arr(B_k)(\rem(B_k) +|f_\tle(B_k)|) & \leq 
\text{~~[by~$(\ddagger)$, $\rem(B_k) + |f_\tle(B_k)| \leq |f_\vtl(B_k)$]}\\
\arr(B_k) |f_\vtl(B_k)|&.
\end{align*}

\noindent
Let us now show that the global time-complexity of  lines~19-21
is in $O(|P_{\mathrm{sim}}||\sra|)$. Let $\max(\cB)=\{M_1,...,M_k\}$
be the maximal elements in $\cB$ so that for any $i\neq j$, 
$M_i\cap M_j = \varnothing$, and in turn we have that
$\textstyle\sum_{M_i\in\max(\cB)} \arr(M_i) \leq |\sra|$.
By using the data structures described in Section~\ref{ds}, the test 
$D\in \Rel(C)$ at line~21 takes constant time.
  Then, the overall complexity of
lines~19-21 is 
\begin{align*}
\textstyle
\sum_{B_i\in \cB} \arr(B_i)\rem(B_i) & = 
\text{~~~~~[as the $M_i$'s are maximal in $\cB$]}\\
\textstyle
\sum_{M_i\in \max(\cB)} \sum_{D\tle M_i} \arr(D) \rem(D) & \leq 
\text{~~~~~[by property $(*)$ above]}\\
\textstyle
\sum_{M_i\in \max(\cB)} \arr(M_i)|\Psim| &= \\
\textstyle
|\Psim| \sum_{M_i\in\max(\cB)} \arr(M_i) & \leq 
\text{~~~~~[as $\textstyle\sum_{M_i\in\max(\cB)} \arr(M_i) \leq |\sra|$]}\\
|\Psim||\sra|&.
\end{align*}

\item[{\rm (H)}] The overall time complexity of lines~22-24 is in
$O(|P_{\mathrm{sim}}||\sra|)$. 

\noindent
\textit{Proof.} 
Let $\cP$ denote the multiset of pairs of blocks 
$(C,D)\in P_i$ that are 
scanned at lines~19-20 at some iteration $i\in \It$
such that $D\in \Rel_i(C)$.
By using the data structures described in Section~\ref{ds}, the test 
$s\not\in \pre(\cup\!\Rel(C))$ and the statement 
$\Rel(C):= \Rel(C)\smallsetminus\{D\}$ take constant time. Moreover, 
the statement $\Remove(C):=\Remove(C)\cup \{s\}$ also
takes constant time because if a state $s$ is added to $\Remove(C)$ at
line~24
then $s$ was not already in $\Remove(C)$ so that this operation 
can be
implemented simply by appending $s$ to the list of states that
represents $\Remove(C)$.  
Therefore, the overall time complexity of the body of the if-then statement
at lines~21-24 is
$\textstyle \sum_{(C,D)\in\cP}\arr(D)$.
We notice the following fact. 
Let $i,j\in \It$ such that $i<j$ and let 
$(C,D_i)$ and $(C,D_j)$ be 
pairs of blocks scanned at lines~19-20, 
respectively, at iterations $i$ and $j$
such that $D_j\subseteq D_i$. Then, if the test $D_i\in \Rel_i(C)$ is
true at iteration $i$ then the test $D_j\in \Rel_j(C)$ is false at
iteration $j$. This is a consequence of the fact that if $D\in
\Rel_i(C)$ then $D$ is removed from $\Rel_i(C)$ at line~22  and
$\cup\!\Rel_j(C) \subseteq \cup\!\Rel_i(C)$ so that $D\cap
\cup\!\Rel_j(C) = \varnothing$. Hence, if
$(C,D),(C,D')\in \cP$  then $D\cap D'=\varnothing$. We define 
the set $\cC \ud \{ C~|~ \exists D.\: (C,D)\in \cP\}$ and 
given $C\in \cC$,  
the multiset 
$\cD_C \ud \{D~|~ (C,D)\in \cP\}$. Observe that $|\cC|$ is bounded by
the number of blocks that appear in some partition $P_i$, so that 
by point~(B), $|\cC| \leq
2(|\Psim|-|P_{\mathrm{in}}|)+|P_{\mathrm{in}}| \leq
2|\Psim|$. Moreover, 
 the
observation above implies that $\cD_C$ is indeed a set and the blocks in
$\cD_C$ are pairwise disjoint.
Thus,
\begin{align*}
\textstyle
  \sum_{(C,D)\in\cP}\arr(D) &= \\
\textstyle
 \sum_{C \in \cC} \sum_{D\in \cD_C} \arr(D)
  &\leq 
\text{~~~~~[as the blocks in $\cD_C$ are pairwise disjoint]}\\
\textstyle
 \sum_{C \in \cC} |\sra| 
&\leq 
\text{~~~~~[as $|\cC| \leq
  2|\Psim|$]}\\
2|\Psim||\sra|&.
\end{align*}

\end{itemize}

\noindent
{}Summing up, we have shown that the overall
time-complexity of $\SA$ is in
$O(|P_{\mathrm{sim}}||\sra|)$. 

\noindent
The space complexity
is in $O(|\Sigma|\log|\Psim| + |\Psim| + |\Psim|^2 +
|\Psim||\Sigma|\log|\Sigma|)=O(|\Psim||\Sigma|\log|\Sigma|)$ where:
\begin{itemize} 
\item[--] 
The pointers from any state $s\in
\Sigma$ to the block of the current partition that contains $s$ are
stored in 
$O(|\Sigma|\log|\Psim|)$ space.
\item[--] The current partition $P$ is stored in $O(|\Psim|)$ space.
\item[--] The current relation $\Rel$ is stored in $O(|\Psim|^2)$ space.
\item[--] Each block of the current partition stores 
the corresponding remove set in $O(|\Sigma|)$ space 
and the integer array $\mathit{RelCount}$ in
$O(|\Sigma|\log|\Sigma|)$, so that these globally take 
$O(|\Psim||\Sigma|\log|\Sigma|)$ space. 
\qedhere
\end{itemize}
\end{proof}

\begin{figure*}[t]
\begin{center} 
{
\scriptsize
\begin{code}[frame=single]
$\mathit[Initialize]$(PartitionRelation P) {
   forall B in P do { 
      B.Remove = $\pre(\Sigma)\smallsetminus\pre$($\cup${C in P | Rel(B,C)}); 
      forall x in $\Sigma$ do B.RelCount(x) = 0;
   }
   forall B in P do 
      forall y in B do 
         forall x in $\pre$({y}) do
            forall C in P such that Rel(C,B) do C.RelCount(x)++;
}

$\SA$(PartitionRelation P) { 
   $\mathit[Initialize]$(P);
   forall B in P such that (B.Remove $\neq \varnothing$) do {
      Set Remove = B.Remove;
      B.Remove = $\varnothing$;
      Set B$_\mathrm[prev]$ = B;
      $\mathit[SplittingProcedure]$(P,Remove);
      ListOfBlocks RemoveList = {D $\in$ P | D $\subseteq$ Remove}; 
      forall C in P such that (C $\cap$ $\pre$(B$_\mathrm[prev]$) $\neq \varnothing$) do
	 forall D in RemoveList do 
	    if (Rel(C,D)) then { 
               Rel(C,D) = 0;
               forall d in D do 
	          forall x in $\pre$(d) do { 
                     C.RelCount(x)--;
		     if (C.RelCount(x) == 0) then {
                        C.Remove = C.Remove $\cup$ {x};
                        P.moveAtTheEnd(C);
                     }
                  }
            }  
   }
}
\end{code}
}
\end{center}
\caption{C++ Pseudocode Implementation of $\SA$.}
\label{fig:sim}
\end{figure*}

\section{Experimental Evaluation}
A pseudocode implementation of the algorithm $\SA$ that shows how the
data structures in Section~\ref{ds} are actually used is in
Figure~\ref{fig:sim}, where 
$\mathit{SplittingProcedure}$ has been introduced above 
in Figure~\ref{fi}. 
We implemented in C++ both our simulation algorithm 
$\SA$ and the $\HHK$ algorithm in order to 
experimentally compare
the time and space performances of $\SA$ and $\HHK$. 
In order to make the comparison as meaningful as possible, these two
C++ implementations use the same data structures for 
storing transitions systems, sets of states and tables. 

Our benchmarks include systems from 
the VLTS (Very Large Transition Systems)
benchmark suite~\cite{vasy}  and some publicly available 
Esterel programs. 
These models are represented as labeled transition systems (LTSs) where labels are
attached to transitions.  Since the versions of $\SA$ and $\HHK$
considered in this paper both
need as input a Kripke
structure, namely a transition system where labels are attached to
states, we exploited a procedure by
Dovier et al.~\cite{dpp04} that transforms a LTS $M$ into
a Kripke structure $M'$ in such a way that bisimulation and simulation
equivalences
on $M$  and $M'$ coincide. This transformation acts as follows:
any labeled transition $\ok{s_1\xrightarrow{l}s_2}$ is replaced by two
unlabeled transitions 
$s_1\ra n$ and $n\ra s_2$, where $n$ is a new node that is labeled
with $l$, while all the original states in $M$ have the same label. 
This labeling provides an initial
partition on $M'$ which is denoted by $P_{\mathrm{in}}$.    
Hence, this transformation grows the size of the model as follows: the number of
transitions is doubled and the number of states of $M'$ is the sum 
of the number of states and transitions of $M$.  
Also, the models cwi\_3\_14, vasy\_5\_9, vasy\_25\_25 and vasy\_8\_38 have
non total transition relations. 
The vasy\_* and cwi\_* systems are
taken from the VLTS suite, while the remaining systems 
are the following Esterel programs: WristWatch and  ShockDance are taken 
from the programming examples of Esterel~\cite{esterel}, 
ObsArbitrer4 and AtLeastOneAck4 are described in the technical 
report~\cite{bouali97}, 
lift, NoAckWithoutReq and one\_pump are provided together with 
the fc2symbmin tool that is used by Xeve, a graphical verification 
environment for Esterel programs~\cite{bouali98,xeve}. 

Our experimental evaluation was carried out
on an Intel Core 2 Duo  1.86 GHz PC, with 2 GB RAM, running Linux
and GNU g++ 4.  The results are summarised in
Table~\ref{results}, where we list the name of the transition
system, the number of states and transitions of the
transformed transition system, the number of blocks of the initial
partition, the number of blocks of the final simulation equivalence partition (that
is known when one algorithm terminates), 
the execution time in
seconds and the allocated memory in MB (this has been obtained by
means of glibc-memusage) both for $\HHK$ and $\SA$, where 
o.o.m.\ means that the algorithm ran out of memory (2GB).

The comparative experimental evaluation shows that $\SA$ outperforms
$\HHK$ both in time and in space.  In fact, the 
experiments demonstrate that $\SA$
improves on $\HHK$ of about two orders of magnitude in time and of one order
of magnitude in space.  The sum of time and space measures on
the eight models where both $\HHK$ and $\SA$ terminate is 64.555 vs.\
1.39 seconds in time and 681.303 vs.\ 52.102 MB in space.  Our
experiments considered 18 models: 
$\HHK$ terminates on 8 models while $\SA$ terminates on 14 of
these 18 models. Also,  the
size of models (states plus transitions) where $\SA$ terminates w.r.t.\
$\HHK$ grows about one order of magnitude.

\begin{table}
  \centering
  \begin{tabular}{|l |l l l |l |l l |l l|}
    \hline
  
    &\multicolumn{3}{|c|}{Input}&Output& \multicolumn{2}{|c|}{$\HHK$} & \multicolumn{2}{c|}{$\SA$} \\ \hline
    Model& $|\Sigma|$ & $|\sra|$ & $|P_{\mathrm{in}}|$ & $|P_{\mathrm{sim}}|$ & 
    Time &Space &
    Time &Space \\
    \hline \hline
    cwi\_1\_2 & 4339 & 4774 & 27 & 2401 & 22.761 & 191& 0.76 & 41\\
    \hline
    cwi\_3\_14 & 18548&29104&  3 &123 & -- & o.o.m.\ & 0.96 & 9\\
    \hline
    vasy\_0\_1 & 1513 &2448&  3 &21 &1.303 & 27& 0.03& 0.229\\
    \hline
    vasy\_10\_56 & 67005&112312&13& ?? & --  & o.o.m.\ & -- & o.o.m.\ \\
    \hline
    vasy\_1\_4 & 5647&8928&7&87 & 37.14 & 407 & 0.28& 2\\
    \hline
    vasy\_18\_73 & 91789&146086&18& ?? & -- & o.o.m.\ & -- & o.o.m.\ \\
    \hline
    vasy\_25\_25 & 50433&50432&25217& ??& -- & o.o.m.\ & -- & o.o.m.\ \\
    \hline
    vasy\_40\_60 & 100013&120014&4& ??  & -- & o.o.m.\ & -- & o.o.m.\ \\
    \hline
    vasy\_5\_9 & 15162&19352&32& 409 & -- & o.o.m.\ & 1.63 & 24 \\
    \hline
    vasy\_8\_24& 33290&48822&12& 1423 & -- & o.o.m.\  & 5.95 & 182 \\
    \hline
    vasy\_8\_38& 47345&76848&82& 963&  -- & o.o.m.\  & 8.15 & 176 \\
    \hline
    WristWatch & 1453 & 1685 & 23 & 1146 & 1.425 & 31 & 0.15 & 6\\
    \hline
	ShockDance & 379 & 459 & 10 & 327 & 0.75 & 2 & 0.03& 0.547\\
    \hline
	ObsArbitrer4 & 17389 & 21394 & 10 & 159 & -- & o.o.m.\ & 0.3 & 11\\
    \hline
	AtLeastOneAck4 & 435 & 507 & 18 & 112 &0.363 & 2 & 0.02& 0.219\\
    \hline
	lift &138  & 163 & 33 & 112 & 0.11 & 0.303 & 0.02 & 0.107\\
    \hline
	NoAckWithoutReq & 1212 & 1372 & 18 & 413 & 0.703 & 21 & 0.1& 2\\
    \hline
	one\_pump & 15774 & 17926 & 22 & 3193 & -- & o.o.m.\ & 13.64 & 194\\
    \hline
  \end{tabular} 
  \caption{Results of the experimental evaluation.}
  \label{results}
\end{table}

\section{Conclusion}

We presented a new efficient algorithm for computing the simulation
preorder in $O(|P_{\mathrm{sim}}||\sra|)$-time and
$O(|P_{\mathrm{sim}}| |\Sigma|\log|\Sigma|)$-space, where
$P_{\mathrm{sim}}$ is the partition induced by simulation equivalence
on some Kripke structure $(\Sigma,\sra)$. This improves the best
available time bound $O(|\Sigma||\sra|)$ given by Henzinger, Henzinger
and Kopke's~\cite{hhk95} and by Bloom and Paige's~\cite{bp95}
simulation algorithms that however suffer from a space complexity that
is bounded from below by $\Omega(|\Sigma|^2)$.  A better space bound
is given by Gentilini et al.'s~\cite{gpp03} algorithm~---~subsequently
corrected by van Glabbeek and Ploeger~\cite{GP08}~---~whose space
complexity is in $O(|P_{\mathrm{sim}}|^2 + |\Sigma| \log
|P_{\mathrm{sim}}|)$, but that runs in $O(|P_{\mathrm{sim}}|^2
|\sra|)$-time. Our algorithm is designed as an adaptation of Henzinger
et al.'s procedure and abstract interpretation techniques are used for
proving its correctness.

As future work, we plan to investigate whether the techniques used for
designing this new simulation algorithm may be generalized and adapted
to other behavioural equivalences like branching simulation
equivalence (a weakening of branching bisimulation equivalence 
\cite{dnv95}).  
It is also interesting to investigate
whether this new algorithm may admit a symbolic version based on BDDs.

\bigskip
\noindent
\textit{Acknowledgements.} 
The authors are grateful to the anonymous referees for their detailed and helpful 
comments and to Silvia Crafa for many useful discussions. 
This work was partially supported by the FIRB Project
``Abstract interpretation and model checking for the verification of
embedded systems'', by the PRIN 2007 Project ``AIDA2007: Abstract
Interpretation Design and Applications'' and by the University of
Padova under the Project ``Formal
methods for specifying and verifying behavioural properties of
software systems''.
This paper is an extended and revised version of~\cite{rt07b}.

\end{document}